\documentclass[fleqn,usenatbib]{mnras}

\usepackage{newtxtext,newtxmath}

\usepackage[T1]{fontenc}

\DeclareRobustCommand{\VAN}[3]{#2}
\let\VANthebibliography\thebibliography
\def\thebibliography{\DeclareRobustCommand{\VAN}[3]{##3}\VANthebibliography}

\def\Omegab{\Omega_\mathrm{b}}

\usepackage{graphicx}	
\usepackage{amsmath}	
\usepackage{enumitem}

\title[Halo substructure generated by bar resonances]{Stellar halo substructure generated by bar resonances}

\author[A. M. Dillamore et al.]{
Adam M. Dillamore,$^{1}$\thanks{E-mail: amd206@cam.ac.uk (AMD)}
Vasily Belokurov,$^{1}$
N. Wyn Evans$^{1}$
and Elliot Y. Davies$^{1}$
\\
$^{1}$Institute of Astronomy, University of Cambridge, Madingley Road, Cambridge CB3 0HA, UK\\
}

\date{Accepted XXX. Received YYY; in original form ZZZ}

\pubyear{2023}

\begin{document}
\label{firstpage}
\pagerange{\pageref{firstpage}--\pageref{lastpage}}
\maketitle

\begin{abstract}
Using data from the {\it Gaia} satellite's Radial Velocity Spectrometer Data Release 3 (RVS, DR3), we find a new and robust feature in the phase space distribution of halo stars. It is a prominent ridge at constant energy and with angular momentum $L_z>0$.
We run test particle simulations of a stellar halo-like distribution of particles in a realistic Milky Way potential with a rotating bar. 
We observe similar structures generated in the simulations from the trapping of particles in resonances with the bar, particularly at the corotation resonance.  
Many of the orbits trapped at the resonances are halo-like, with large vertical excursions from the disc.
The location of the observed structure in energy space is consistent with a bar pattern speed in the range $\Omega_\mathrm{b}\approx35-40$~km\,s$^{-1}$\,kpc$^{-1}$.
Overall, the effect of the resonances is to give the inner stellar halo a mild, net spin in the direction of the bar's rotation. As the distribution of the angular momentum becomes asymmetric, a population of stars with positive mean $L_z$ and low vertical action is created.
The variation of the average rotational velocity of the simulated stellar halo with radius is similar to the behaviour of metal-poor stars in data from the APOGEE survey.
Though the effects of bar resonances have long been known in the Galactic disc, this is strong evidence that the bar can drive changes even in the diffuse and extended stellar halo through its resonances.
\end{abstract}




\section{Introduction}

The first person to entertain the idea that the Milky Way is a barred galaxy seems to have been \citet{Va64}, based on kinematic evidence of deviations from circular motions in the 21-cm line profiles near the Galactic centre. Curiously, though, \citet{Va64} has the orientation of the bar wrong, with the near side at negative Galactic longitudes, rather than positive. The first clear evidence that the Milky Way is barred with the near side at positive longitudes seems to have come from infrared photometry. Both integrated 2.4 $\mu$m emission~\citep{Bl91} and the distribution of IRAS Miras~\citep{Wh92} clearly showed the inner Galaxy is brighter at positive longitudes. Work on the gas kinematics~\citep{Bi91}, the starcounts of red clump giants~\citep{St94} and especially the infrared photometric maps obtained by the COBE satellite~\citep{We94} added to the growing evidence of a barred Milky Way, which became a concensus by the mid 1990s.

Many of the parameters of the bar in the Milky Way have been constrained in recent years. The pattern speed of the bar $\Omegab$ controls its length. The stellar orbits that support the bar cannot exist much beyond corotation~\citep{Co80}. Older studies matching the gas flows suggested a fast and short bar with $\Omegab$ in the range 50 to 60 kms$^{-1}$ kpc$^{-1}$ ~\citep{Fu99,Bi03}. As the quality and quantity of the gas data has improved, more recent works have determined lower values, consistent with a slower and longer bar~\citep{So15,Li22}. The pattern speed of the bar is now thought to lie in the range 35 to 40 kms$^{-1}$ kpc$^{-1}$ \citep[e.g.][]{Po17,wang2013,Sa19,binney2020,chiba2021_treering}. This is consistent with results from stellar kinematics -- for example, use of the projected continuity equation or \citet{Tr84} method by \citet{Sa19}, or modelling of multiple bulge stellar populations using spectroscopic data~\citep{Po17}. 

The bar angle -- that is the angle between the line joining the Sun and Galactic Centre  and the major axis of the bar -- is more uncertain. Studies reaching to longitudes $|\ell| > 10^\circ$ often find bar angles $\approx 45^\circ$ ~\citep[e.g.,][]{Go12}. In studies confined to the inner parts, bar angles of $20^\circ$-$35^\circ$ are typical~\citep[e.g.,][]{St97,wegg2015,Simion2017}. It has been suggested that the Galaxy contains two bars, with the central bar not aligned with the long bar \citep[e.g.][]{hammersley2000,benjamin2005,cabrera-lavers2007,cabrera-lavers2008,churchwell2009}. Such a configuration is however difficult to explain dynamically, and it has been shown that these observations can be reproduced with a single structure made up of a boxy bulge and long bar \citep{martinez-valpuesta2011}.

Resonant perturbation theory \citep{LBK, Ly73} was introduced into galactic dynamics to understand the trapping of stellar orbits by bar-like perturbations. For nearly circular orbits in a disc, the effects of the bar are most significant at the corotation and Lindblad resonances~\citep[e.g.,][]{Shu}. \citet{Ka91} first suggested that this effect may be responsible for creation of the Hyades and Sirius streams in the disc. The closed orbits inside and outside the outer Lindblad Resonance are elongated perpendicular and parallel to the bar in this picture. Subsequently, \citet{De98} used \textit{HIPPARCOS} data to demonstrate the existence of abundant substructure or moving groups in the disc, which he ascribed partly to resonance effects. The \textit{Gaia} data releases have revealed abundant ridges, undulations and streams in velocity or action space, sculpted by the Galactic bar~\citep[e.g.,][]{Fr19,
khoperskov2020,trick2021,Tr22,wheeler2022}.

The idea that the Galactic bar can create substructure in the disc is well-established. The disc orbits are nearly circular and confined largely to the Galactic plane, so it is easy to see how a rotating bisymmetric disturbance can couple to the orbits. By contrast, in the halo, stars are moving on eccentric orbits and can reach heights of many kiloparsecs above the Galactic plane. While less well studied, numerical experiments suggest that these stellar orbits may still be susceptible to trapping by bar resonances ~\citep{moreno2015}. In addition, the effects of bars on the dark halo has been the subject of much debate over the last decades~\citep[e.g.,][]{Ce07, Deb98,De00, We02}. The slowing of bars by dynamical friction of the dark matter halo can change the central parts from cusps to cores \citep{We02}. This induces profound changes in the orbits of the dark matter particles in the halo \citep{athanassoula2002,collier2019}.

Recent data from the third data release (DR3) of \textit{Gaia} \citep{gaia, gaia_dr3} has revealed previously undetected substructure in the stellar halo of the Milky Way. By plotting Galactocentric radial velocity $v_r$ against Galactocentric radius $r$, \citet{belokurov_chevrons} showed that there are multiple `chevron'-shaped overdensities in this radial phase space. These bear a close resemblance to structures which result from the phase mixing of debris from a merging satellite on an eccentric or radial orbit \citep{filmore1984,dong-paez2022,davies2023_ironing}. \citet{belokurov_chevrons} showed that they are present at both low and high metallicity, and are still visible when only stars with [Fe/H] $>-0.7$ are included. However, stars originating from the Milky Way's accreted satellites have lower metallicity than this. The \textit{Gaia} Sausage-Enceladus \citep{belokurov2018,helmi2018} was one of the most massive mergers in the Milky Way's history, and its comparatively high metallicity debris dominates the stellar halo in the solar neighbourhood \citet{naidu2020}. However, only a very small proportion of its stars have [Fe/H] $>-0.7$ \citep{naidu2020,feuillet2021}. This challenges the assumption that the radial phase space structures are due to accreted material.

There are already indications that bar-driven features in the stellar halo are possible. For example, motivated by the shortness of the Ophiuchus Stream, \citet{Ha16} used numerical simulations to show that its properties may have been influenced by its interaction with the bar, despite the Stream stars lying at heights of $\approx 5$ kpc above the Galactic plane. \citet{Sc19} examined the halo moving groups G18-39 and G21-22 as possible resonant structures generated by the bar -- albeit with a pattern speed 45-55 kms$^{-1}$kpc$^{-1}$ which is rather high nowadays. These mildly retrograde moving groups were discovered in \citet{Si12}, who suggested that they may be debris from the unusual and retrograde globular cluster $\omega$ Centauri. \citet{My18} identified the resonance generating the Hercules Stream in the solar neighbourhood as present even in stars with metallicities as low as [Fe/H] $\approx$ -2.9 and so normally associated with thick disc and halo. Very recently, the interaction of a rotating bar with radial phase space chevrons has been studied by \citet{davies2023_bar}. They showed that with realistic values of the pattern speed, the bar is capable of blurring and destroying much of the phase space structure resulting from a satellite merger. If a bar can destroy, then it is natural to ask whether it can also create substructure.

In this work, we examine the creation of substructure from a smooth stellar halo via resonances. We identify a prominent ridge in energy and angular momentum space (a proxy for action space) and show that such a feature is a natural outcome of bar-driven resonances in the stellar halo. The prominence and narrowness of this ridge provides a novel method of measuring the pattern speed of the bar.

The paper is arranged as follows. In Section \ref{section:dynamics}, we summarize the dynamics of orbits in rotating potentials and introduce the principal bar resonances. We use data from \textit{Gaia} DR3 in Section \ref{section:data} to reveal structures in energy-angular momentum and radial phase space. In Section \ref{section:simulations}, we describe our simulations of the stellar halo and bar, and compare them to the data. Finally we present our conclusions in Section \ref{section:conclusions}.

\section{Dynamics in a rotating potential}
\label{section:dynamics}

Consider a steady non-axisymmetric potential rotating with pattern speed (angular frequency) $\Omega_\mathrm{b}$ about the $z$-axis. Although the energy $E$ and $z$ component of angular momentum $L_z$ of a particle can vary, a linear combination of the two known as the Jacobi integral is conserved~\citep{binney_tremaine}. This is defined as
\begin{equation}
    H_\mathrm{J}=E-\Omega_\mathrm{b}L_z.
\end{equation}
The Jacobi integral is the energy in the rotating frame. As the potential is steady in this frame, it follows from time-invariance that the Jacobi integral is conserved. Hence in the $E$ versus $L_z$ plane, stars are constrained to move along straight lines of gradient $\Omega_\mathrm{b}$, provided the pattern speed remains steady.

Action-angle coordinates are a set of canonical coordinates useful in nearly integrable systems where most orbits are regular and not chaotic~\citep[e.g.][]{Ar78}. Each star is described by three actions $J_i$ which are constant and three angles $\theta_i$ which increase linearly with frequencies $\Omega_i$. As an example, a slightly eccentric and inclined orbit in a weakly non-axisymmetric potential can be approximated as motion on an epicycle around a guiding centre. This centre follows a circular orbit in the $z=0$ plane with azimuthal frequency $\Omega_\phi$, set by the rotation curve of the potential. The particle oscillates in the (cylindrical polar) $R$-direction with the epicyclic frequency $\Omega_R$ and in the $z$-direction with the vertical frequency $\Omega_z$. Action-angle coordinates can be computed for general orbits in axisymmetric galactic potentials using modern stellar dynamical packages like \textsc{Agama} ~\citep{agama}.

The locations of the principal resonances in space is in general hard to establish. Only for orbits confined to the disc plane are matters reasonably straightforward. In the frame corotating with the bar at frequency equal to the pattern speed $\Omega_\mathrm{b}$, the mean azimuthal frequency of a star is $\Omega_\phi-\Omega_\mathrm{b}$. We define the ratio of frequencies in this frame $r_\Omega\equiv(\Omega_\phi-\Omega_\mathrm{b})/\Omega_R$. An orbit is resonant with the bar if $r_\Omega=\pm n/m$, where $n$ and $m$ are integers. The most important of these are the corotation resonance ($\Omega_\phi=\Omega_\mathrm{b}$, $r_\Omega=0$) and the Lindblad resonances ($r_\Omega=\pm1/m$) \citep{binney_tremaine}. For orbits that explore the stellar halo, the location of the resonances can change with height above or below the Galactic plane \citep{moreno2015,trick2021}.

Why are the resonant orbits so important? Suppose a bar-like disturbance rotating at angular frequency $\Omegab$ is applied to the disc. On each traverse, the resonant stars meet the crests and troughs of the perturbation potential at the same spots in their orbits and this causes secular changes in the orbital elements~\cite[e.g.][]{Ly73, LBK, Co97,Mo15}. The non-resonant stars feel only periodic fluctuations that average to zero in the long term. As the strength of the bar's perturbation increases, stars near the locus of exact resonance are captured into libration around the parent closed periodic orbit. So, the neighbourhoods of the resonances are the regions of a galaxy where a bar can produce long-lived changes in the stellar populations.

\section{Data}
\label{section:data}

\subsection{\textit{Gaia} DR3 RVS sample}

We use the sample of stars from \textit{Gaia} DR3 \citep{gaia,gaia_dr3} described by \citet{belokurov_chevrons} which includes line-of-sight velocity measurements from the Radial Velocity Spectrometer \citep[RVS,][]{gaia_rvs}. We use distances calculated by \citet{bailer-jones2021}. This sample has selection cuts on parallax error ($\varpi/\sigma_\varpi>10$) and heliocentric distance ($D<15$~kpc). All sources within $1.5^\circ$ of known globular clusters closer than 5~kpc have been removed \citep{belokurov_chevrons}. This leaves $\sim25$ million sources out of the $\sim34$ million with line-of-sight velocity measurements. For Section~\ref{section:radial_phase_space} we further limit our sample to stars with metallicity ([Fe/H]) values calculated by \citet{belokurov_chevrons}. These were derived from BP/RP spectra calibrated with data from APOGEE DR17 \citep{apogee_dr17}. See \citet{belokurov_chevrons} for a detailed description. This provides us with a final sample of 22.5 million sources, all with 6D phase space and [Fe/H] measurements.

\begin{figure}
  \centering
  \includegraphics[width=\columnwidth]{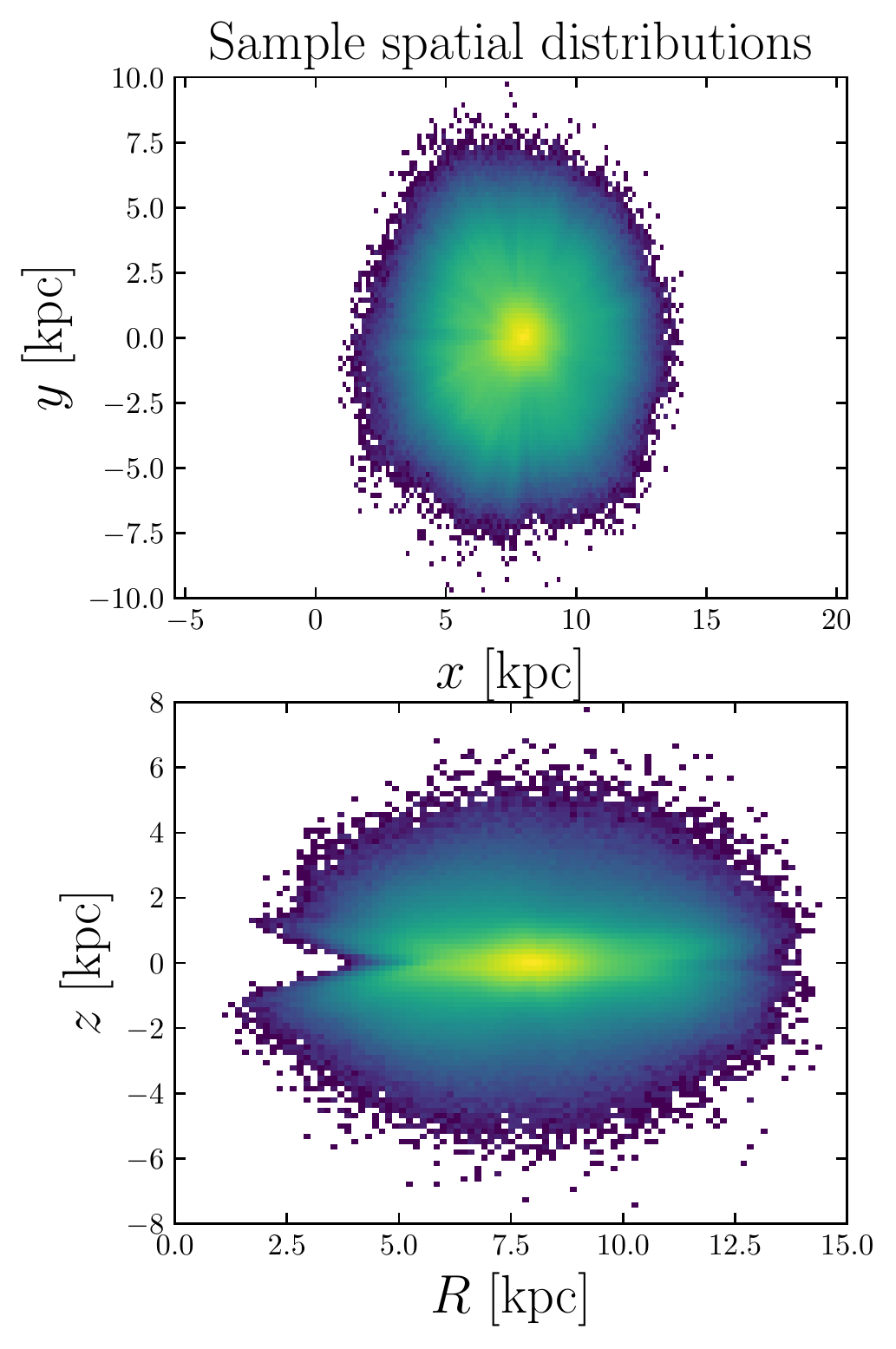}
  \caption{Distributions of our \textit{Gaia} DR3 sample in Galactocentric Cartesian coordinates (top panel) and cylindrical coordinates (bottom panel). The Sun is located at $x=R=8$ kpc, $y=z=0$ kpc.}
   \label{fig:sample_spatial_dists}
\end{figure}

We transform the positions and velocities into a Galactocentric left-handed coordinate system in which the Sun's position and velocity are $(x_\odot,y_\odot,z_\odot)=(8,0,0)$~kpc and $(v_{\odot x},v_{\odot y},v_{\odot z})=(-9.3,251.5,8.59)$~km\,s$^{-1}$ \citep[as reported by][]{gaia_dr3_disc}. The $x$-$y$ and $R$-$z$ distributions of this sample are shown in Fig.~\ref{fig:sample_spatial_dists}.

\begin{figure}
  \centering
  \includegraphics[width=\columnwidth]{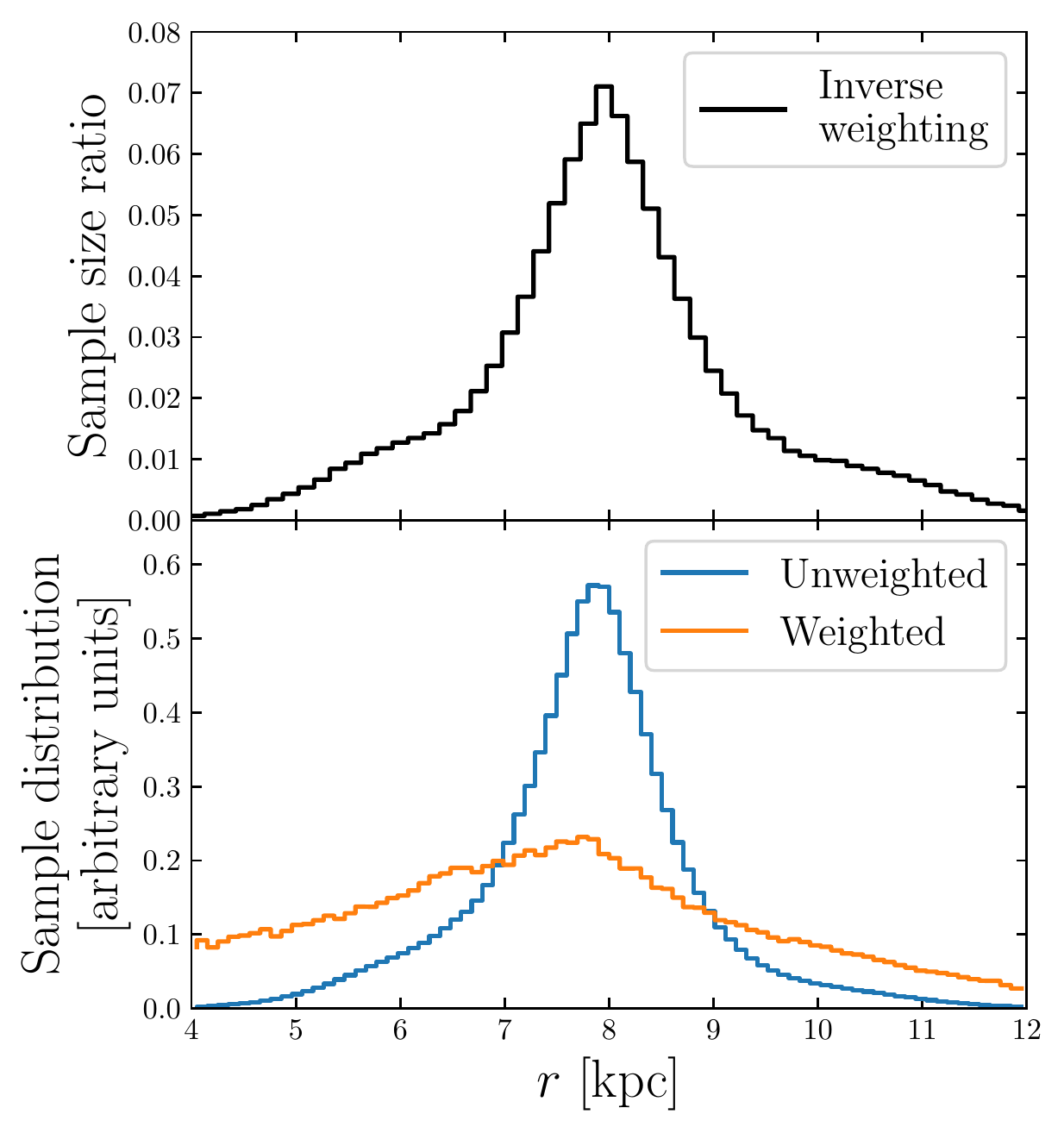}
  \caption{\textbf{Top panel:} The inverse of the weighting function against Galactocentric radius $r$. This is the ratio of the size of the RVS sample with a parallax error cut applied ($\varpi/\sigma_\varpi>10$) to the size of the DR3 sample with photo-geometric distance measurements. There is a strong peak between 7 and 9 kpc due to the high concentration of the former sample around the Sun. \textbf{Bottom panel:} histograms of the unweighted (blue) and weighted (orange) $r$ distributions of our sample. This demonstrates that the weighting largely removes the strong peak around the Sun at $r=8$ kpc.}
   \label{fig:weighting}
\end{figure}

\begin{figure}
  \centering
  \includegraphics[width=\columnwidth]{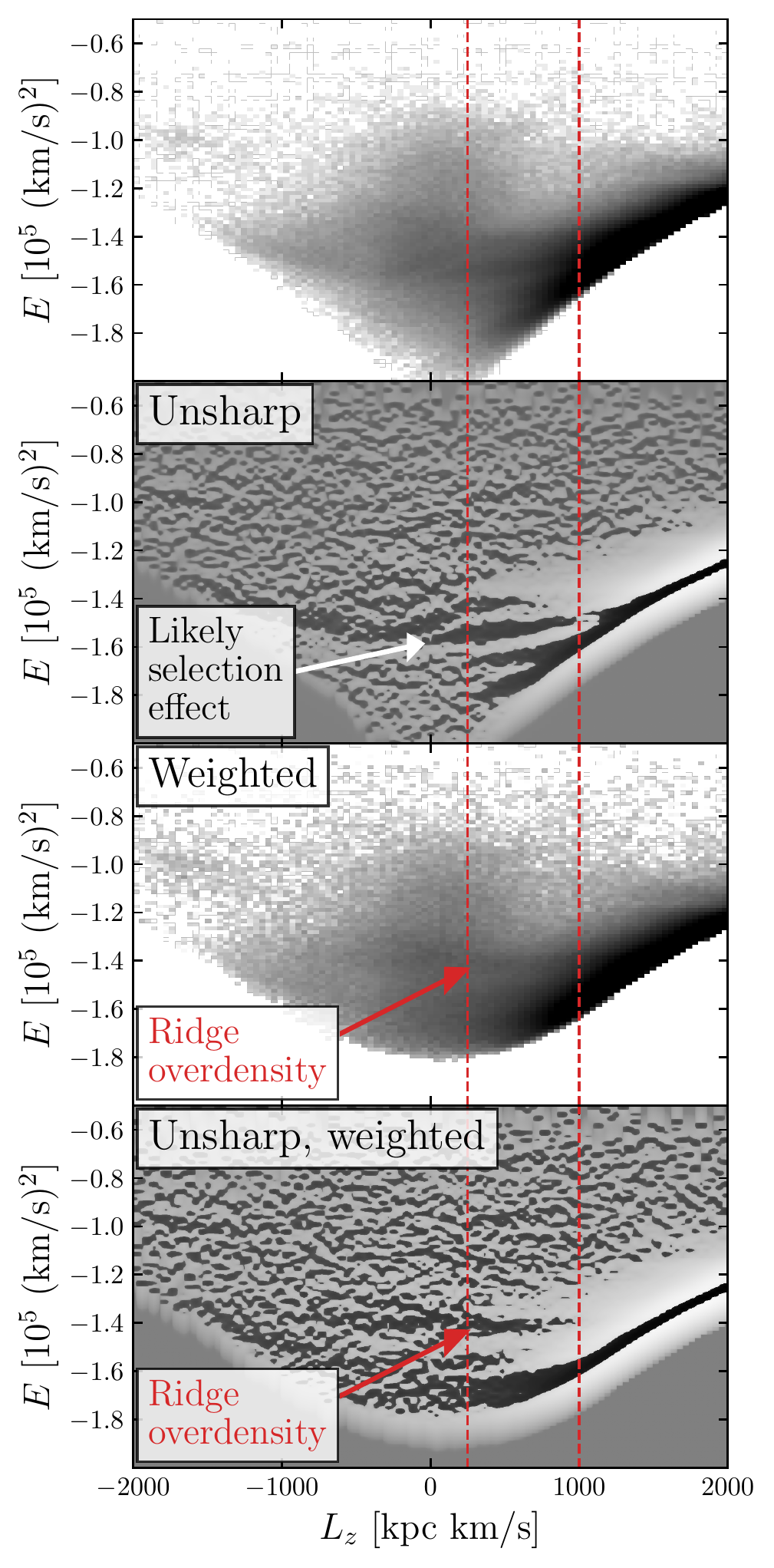}
  \caption{\textbf{Top panel:} distribution of energy $E$ vs angular momentum $L_z$, where prograde orbits have $L_z>0$. The red dashed lines mark the angular momentum cut used in Fig.~\ref{fig:data_chevrons}. The highly populated disc can be seen on the right-hand side of the distribution. \textbf{2nd panel:} as above, but with unsharp filtering applied; a background smoothed with a Gaussian kernel in $E$ has been subtracted off. We use 100 pixels along the $E$-axis, and the Gaussian kernel has a standard deviation of 2 pixels. Black (white) pixels denote overdensities (underdensities). This reveals several ridges at roughly constant energy with mostly $L_z>0$. \textbf{Bottom two panels}: As above, but all data points are weighted based on the number of stars in the RVS sample as a function of radius. This largely removes the ridge at $E\approx-1.5\times10^5$ km$^2$\,s$^{-2}$, which was likely due to a selection effect. The highest energy ridge (at $E\approx-1.4\times10^5$ km$^2$\,s$^{-2}$) remains.}
   \label{fig:E_Lz_data}
\end{figure}

\begin{figure*}
  \centering
  \includegraphics[width=0.8\textwidth]{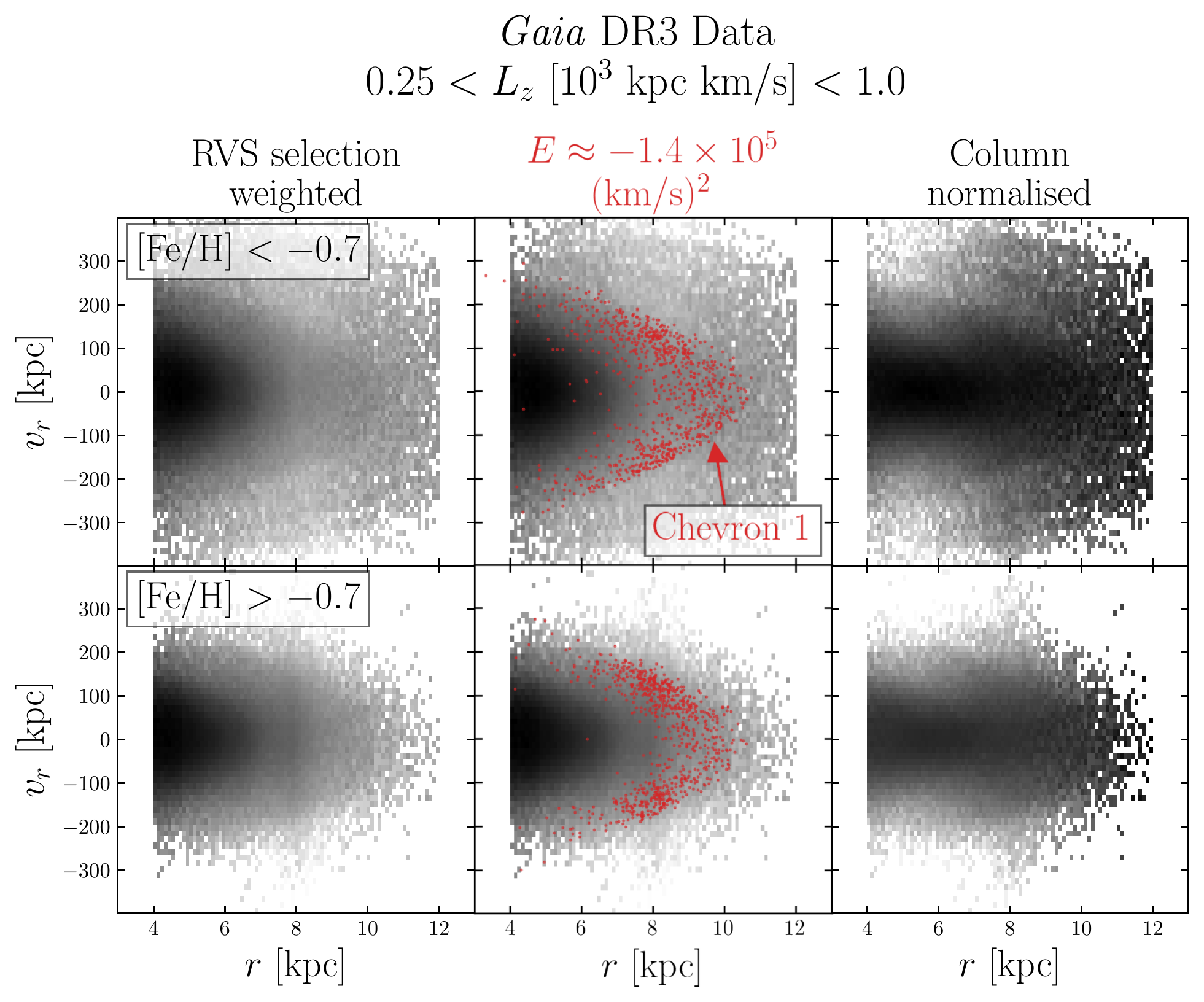}
  \caption{Radial phase space of the DR3 RVS sample, with an $L_z$ cut applied. The top (bottom) row shows low (high) metallicity stars, with the threshold set at [Fe/H] $=-0.7$. The high metallicity sample is dominated by \textit{in situ} stars. In the middle column, we plot the positions of particles with energy $E\approx-1.4\times10^5$ km$^2$\,s$^{-2}$, which corresponds to the horizontal ridge in $E$-$L_z$ space (Fig.~\ref{fig:E_Lz_data}). In the right-hand column, we column-normalise the histograms (i.e. each column of pixels has the same total count). In the top row, several chevron-shaped overdensities are visible, the most prominent of which (peaking at $r\approx10.5$ kpc) is also visible at high metallicity. As shown by the middle column, this corresponds to the energy of the horizontal overdensity in Fig.~\ref{fig:E_Lz_data}. Following \citet{belokurov_chevrons}, we denote this feature `Chevron 1'. The dark region (at $r\lesssim7$ kpc) is due to the disc, and moves depending on the $L_z$ cut.}
   \label{fig:data_chevrons}
\end{figure*}

\begin{figure}
  \centering
  \includegraphics[width=0.9\columnwidth]{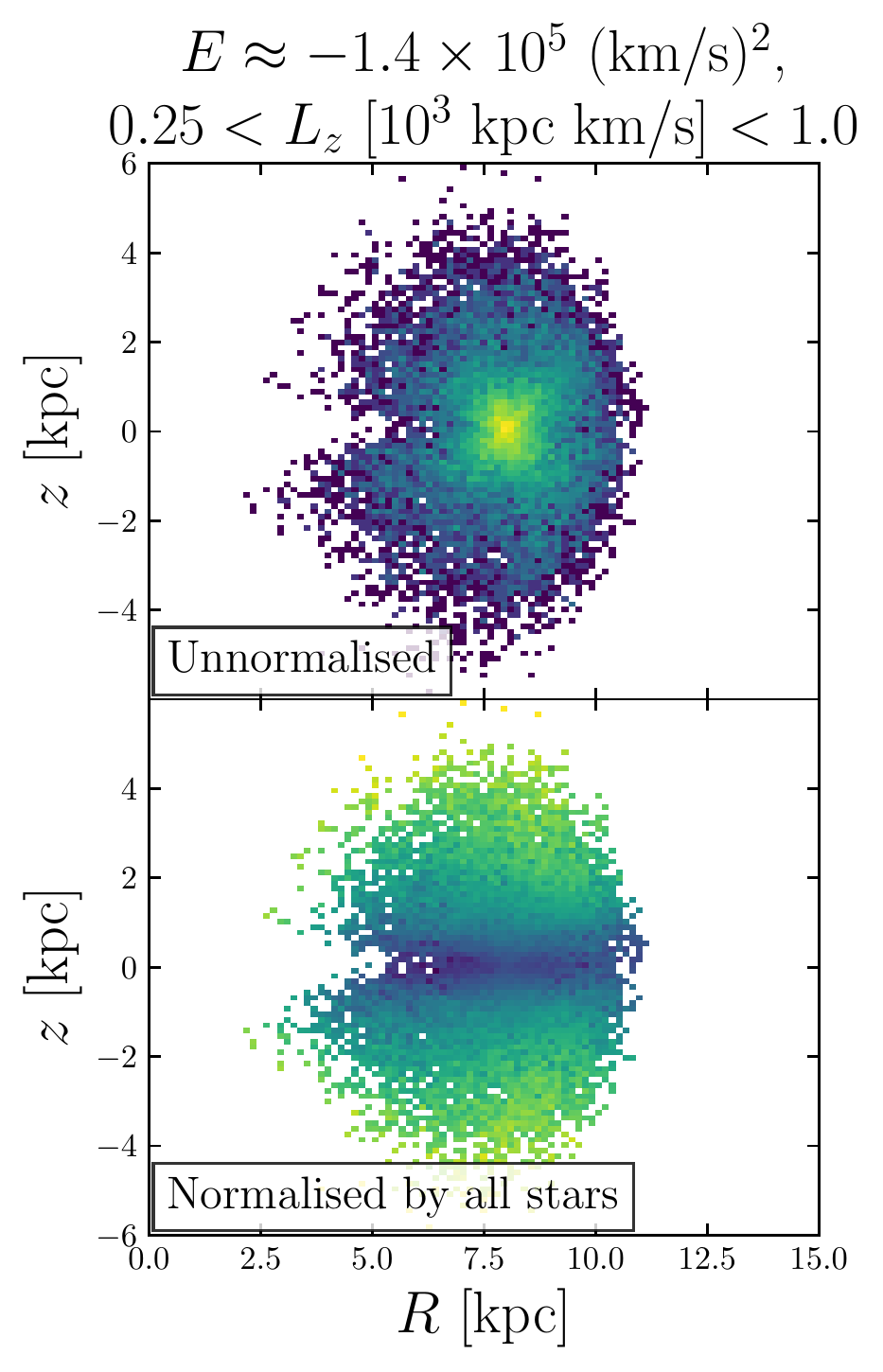}
  \caption{Radial and vertical distribution of stars close to the ridge at $E=-1.4\times10^5$~km$^2$\,s$^{-2}$, within the angular momentum cut $0.25<L_z$ [$10^3$~km\,s$^{-1}$] $<1.0$. In the bottom panel the distribution is normalised by the $R$-$z$ distribution of all stars in the sample. The selected stars are spread over a wide range of $z$, with no large excess close to $z=0$ (other than near the Sun). This is emphasised by the lower panel, which shows the depletion of stars near the disc compared to the sample as a whole. The overdensity therefore has a large contribution from the stellar halo.}
   \label{fig:corotation_R_z}

    \centering
  \includegraphics[width=\columnwidth]{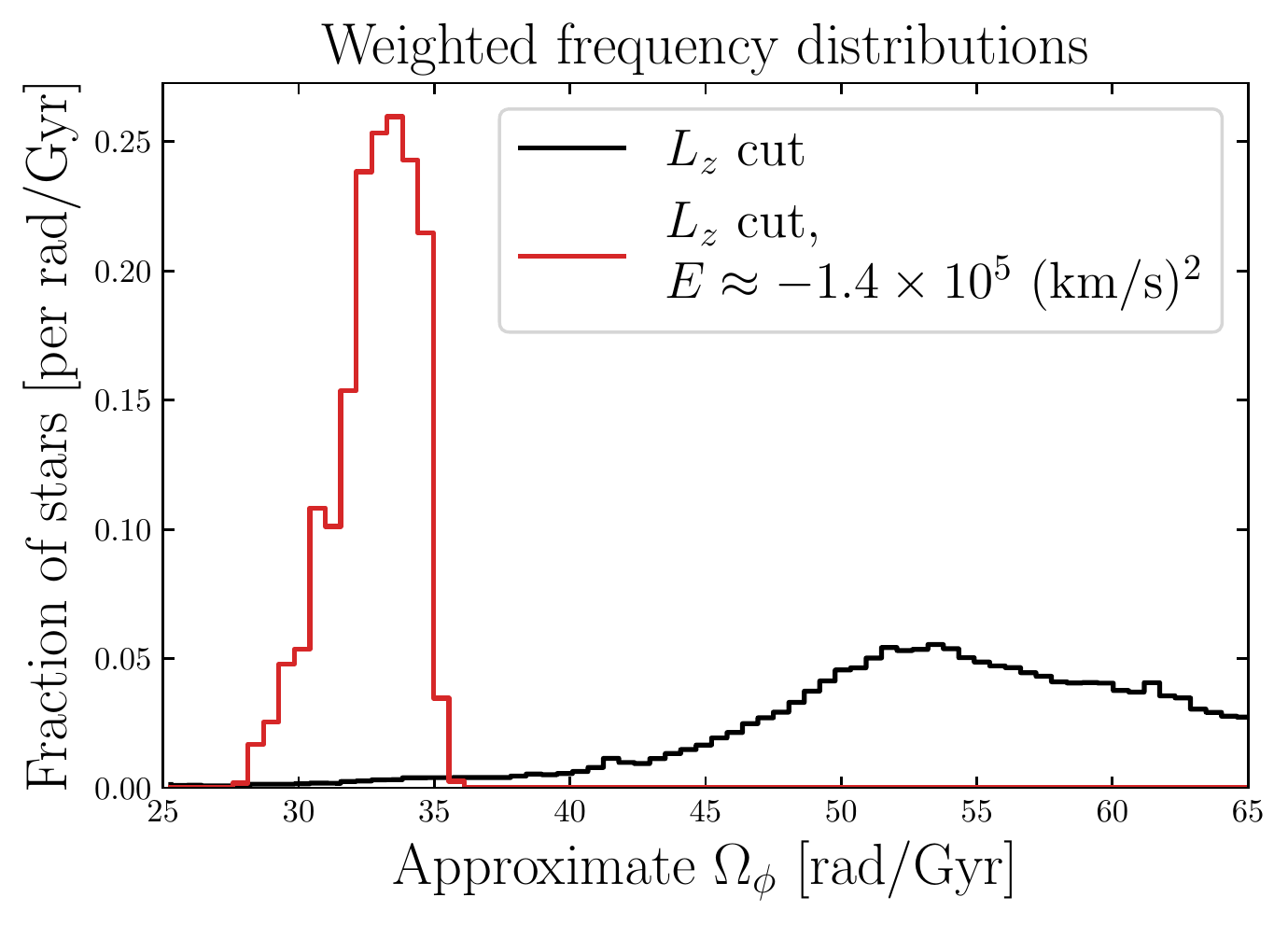}
  \caption{Distributions of the azimuthal frequency $\Omega_\phi$ within the angular momentum cut $0.25<L_z$ [$10^3$~km\,s$^{-1}$] $<1.0$] $<1.0$. The black histogram shows all stars, and the red shows those with energies close to $E=-1.4\times10^5$ km$^2$\,s$^{-2}$. We have applied the weighting discussed in Section~\ref{section:weighting}. Stars close to the energy overdensity have frequencies strongly peaked around 34 rad$\,$Gyr$^{-1}$, similar to some estimates of the bar's pattern speed.}
   \label{fig:data_freqs}
   
\end{figure}

\begin{figure}
  \centering
  \includegraphics[width=0.9\columnwidth]{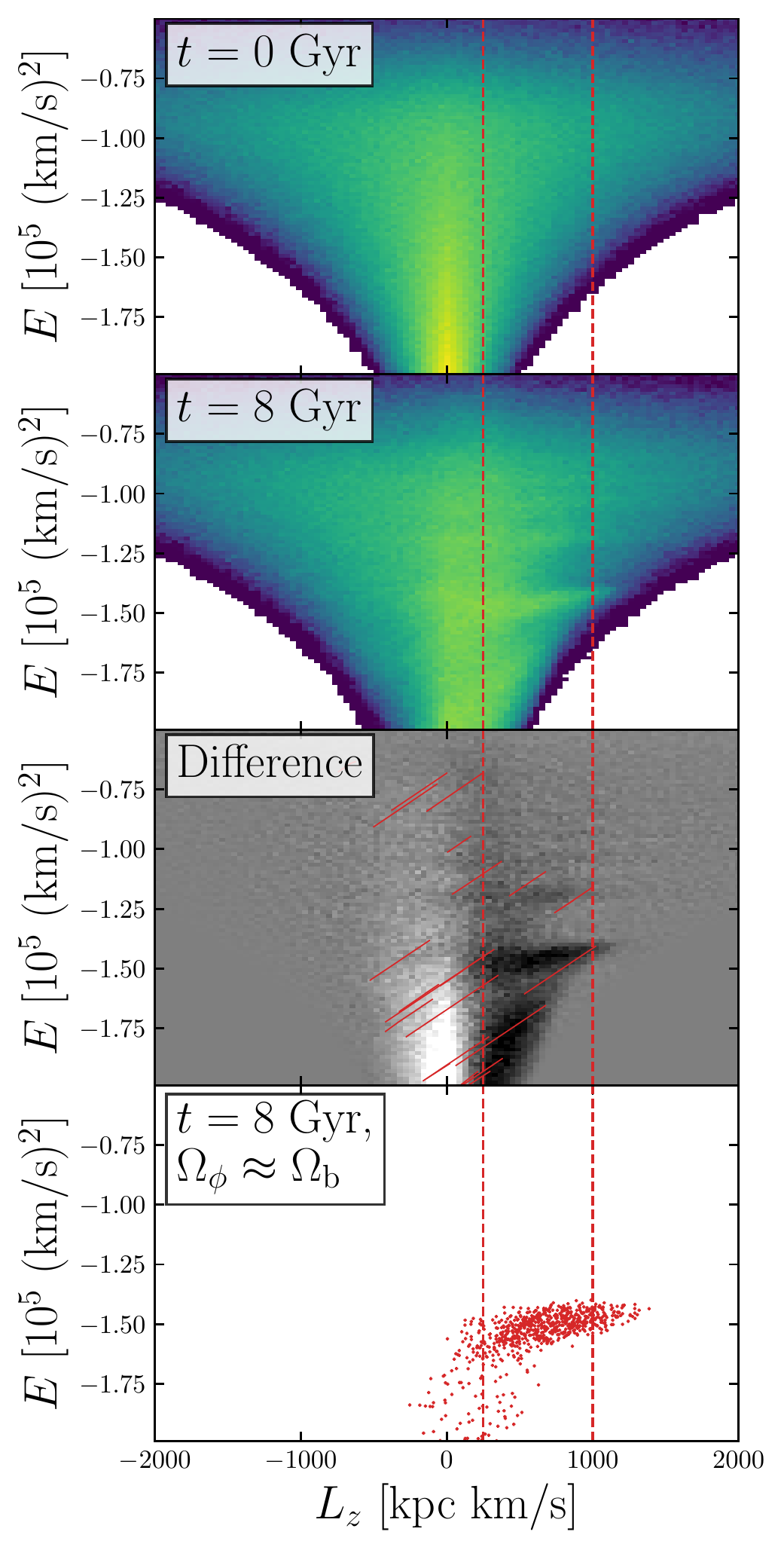}
  \caption{\textbf{Top two panels}: Distributions of $E$ and $L_z$ of the star particles at the beginning and end of the simulation. The colour map uses the same logarithmic normalisation for each panel, and as in Fig.~\ref{fig:E_Lz_data} the red dashed lines mark the $L_z$ cut. After the growth of the bar, ridge-like overdensities at certain energies emerge from the initially smooth distribution. These are at $L_z>0$ (i.e. orbiting in the same direction as the bar's rotation). \textbf{3rd panel}: Difference between the $t=0$ and later distributions. Black (white) pixels mark where the density has increased (decreased). This emphasises that particles have moved away from $L_z\approx0$ and $L_z<0$ into the ridges at $L_z>0$. The red diagonal streaks mark the paths taken by a selection of particles, and have gradients close to the pattern speed. \textbf{Bottom panel:} stars close to the corotation resonance ($\Omega_\phi=\Omega_\mathrm{b}$) at the end of the simulation. These are almost all located within the lowest energy overdensity at $E\approx-1.5\times10^5$ km$^2$\,s$^{-2}$, strongly implying that this overdensity is a manifestation of this resonance.}
   \label{fig:E_Lz_sim}
\end{figure}

\begin{figure*}
  \centering
  \includegraphics[width=0.8\textwidth]{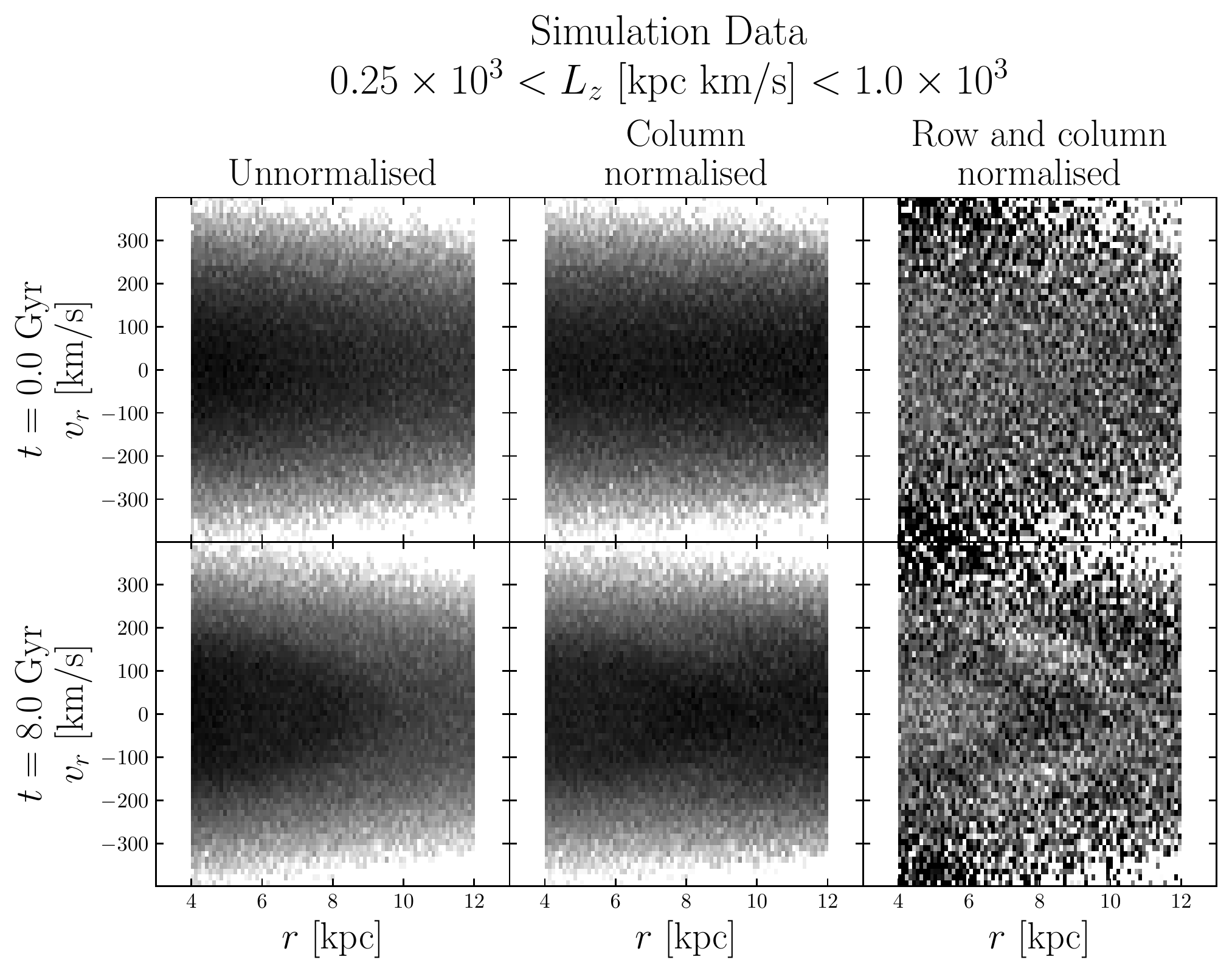}
  \caption{Radial phase space of the simulated stars, with the same $L_z$ cut as used on the data in Fig.~\ref{fig:data_chevrons}. The top and bottom rows show the start and end of the simulation respectively. From left to right, the columns are unnormalised, normalised by column, and normalised by both row and column. The distribution is initially smooth, with no structure visible. However, at the end of the simulation after the growth of the bar, a wedge-shaped overdensity with its tip at $\sim10$ kpc is visible. This is revealed more clearly in the right-hand column, and bears a close resemblance to Chevron 1 seen in the data in Fig.~\ref{fig:data_chevrons}.}
   \label{fig:sim_chevrons}
\end{figure*}

\begin{figure}
  \centering
  \includegraphics[width=0.8\columnwidth]{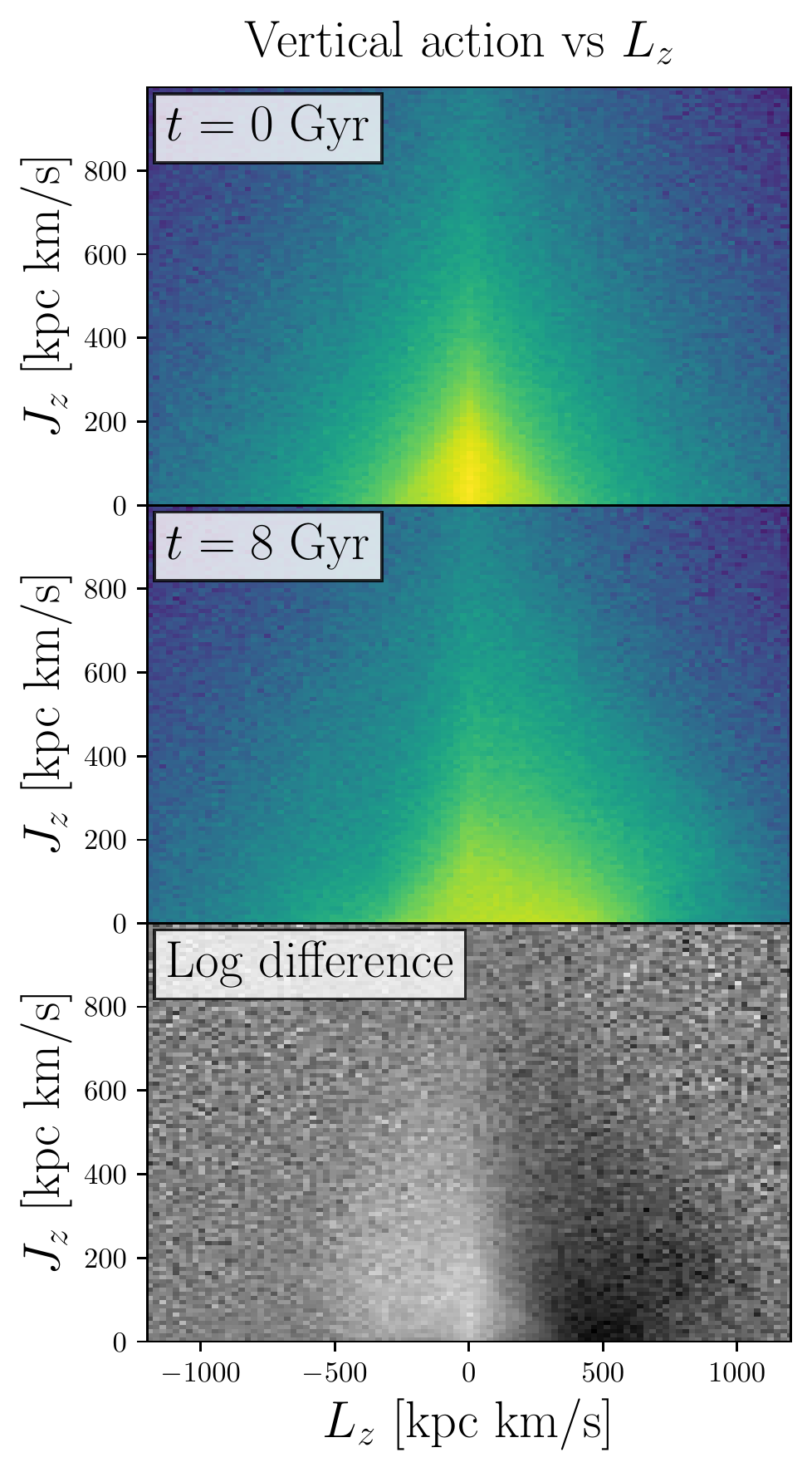}
  \caption{Vertical action $J_z$ vs angular momentum $L_z$ at the beginning and end of the simulation. The bottom panel shows the ratio of the two histograms, where black (white) pixels show where the density has increased (decreased). This indicates that stars at low $J_z$ are more likely to be pushed to high $L_z$ by the bar.}
   \label{fig:Jz_Lz}
\end{figure}

\begin{figure}
  \centering
  \includegraphics[width=\columnwidth]{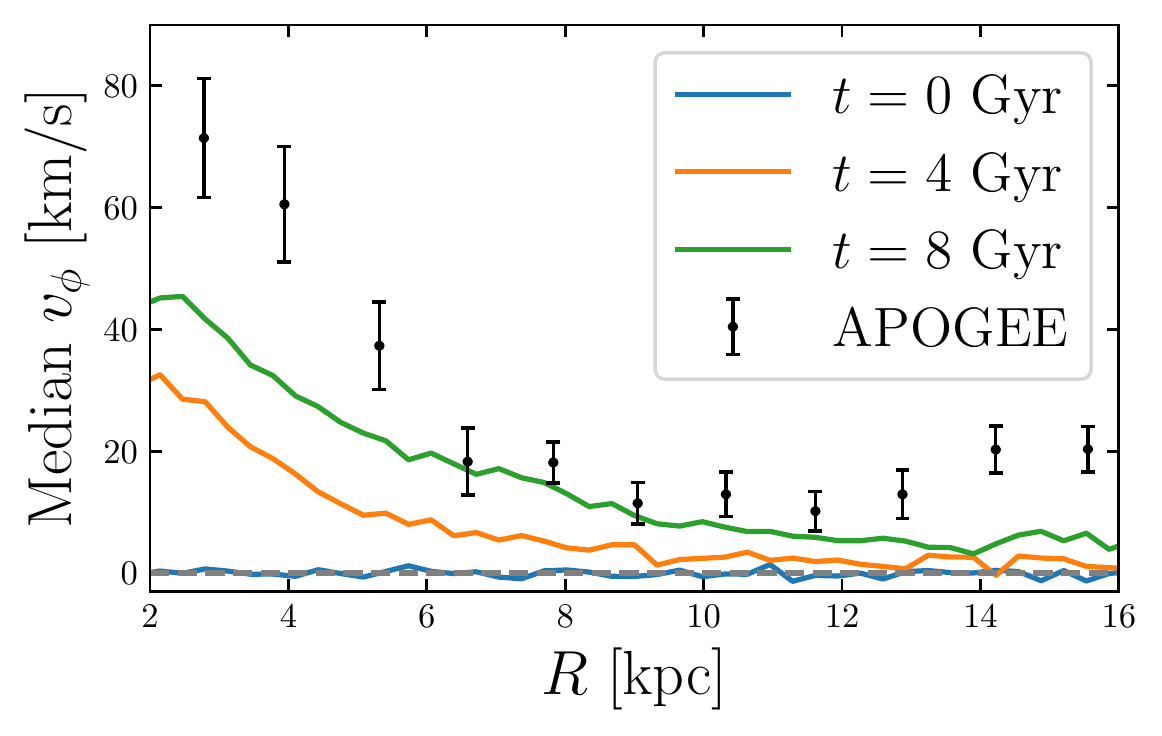}
  \caption{Median azimuthal velocity $\langle v_\phi\rangle$ as a function of cylindrical radius $R$ in each snapshot of the simulations (coloured lines) compared to observed values for low [Fe/H] stars from APOGEE DR17 (black points). While initially the halo has no net spin, after the bar grows the mean azimuthal velocity is positive, particularly inside $8$ kpc. The observed values show a similar trend, albeit with a higher median $v_\phi$.}
   \label{fig:density_profile_change}
\end{figure}


\begin{figure}
  \centering
  \includegraphics[width=\columnwidth]{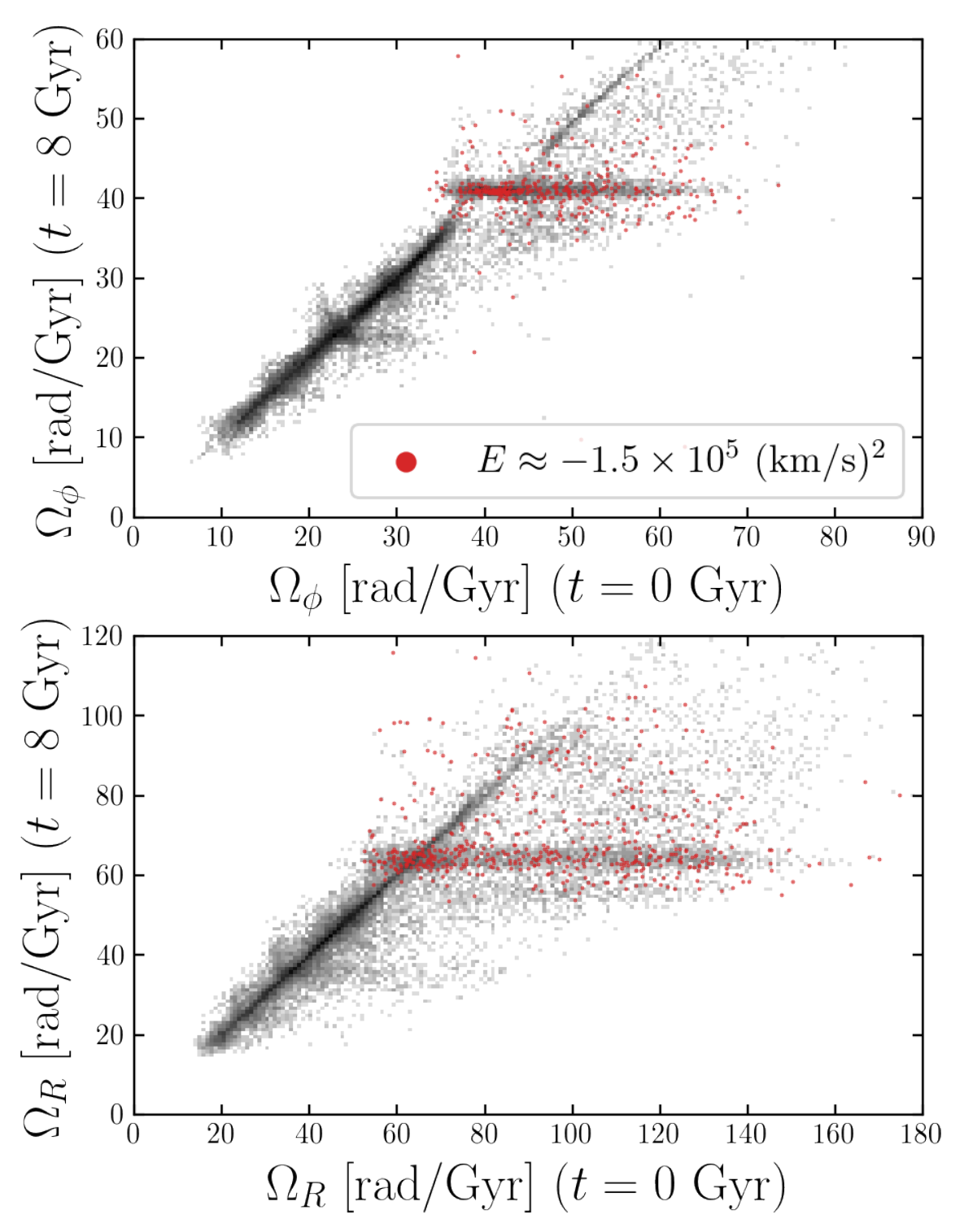}
  \caption{\textbf{Top panel:} initial ($x$-axis) and final ($y$-axis) azimuthal frequencies $\Omega_\phi$ of simulated stars. These are calculated by integrating orbits from the initial and final phase space positions in a potential of constant pattern speed. The red points show stars with final energy close to $E=-1.5\times10^5$ km$^2$\,s$^{-2}$, the approximate energy of the most prominent overdensity in Fig.~\ref{fig:E_Lz_sim}. \textbf{Bottom panel:} as above, but instead showing radial frequencies $\Omega_R$. In both cases horizontal stripes indicate that the final frequencies are highly clustered about certain values. These include $\Omega_\phi=40$ km\,s$^{-1}$\,kpc$^{-1}$, the bar pattern speed. Most of the red points lie close to this frequency, showing that this corresponds to the prominent energy overdensity in Fig.~\ref{fig:E_Lz_sim}.}
   \label{fig:Omega_sim}
\end{figure}

\begin{figure}
  \centering
  \includegraphics[width=\columnwidth]{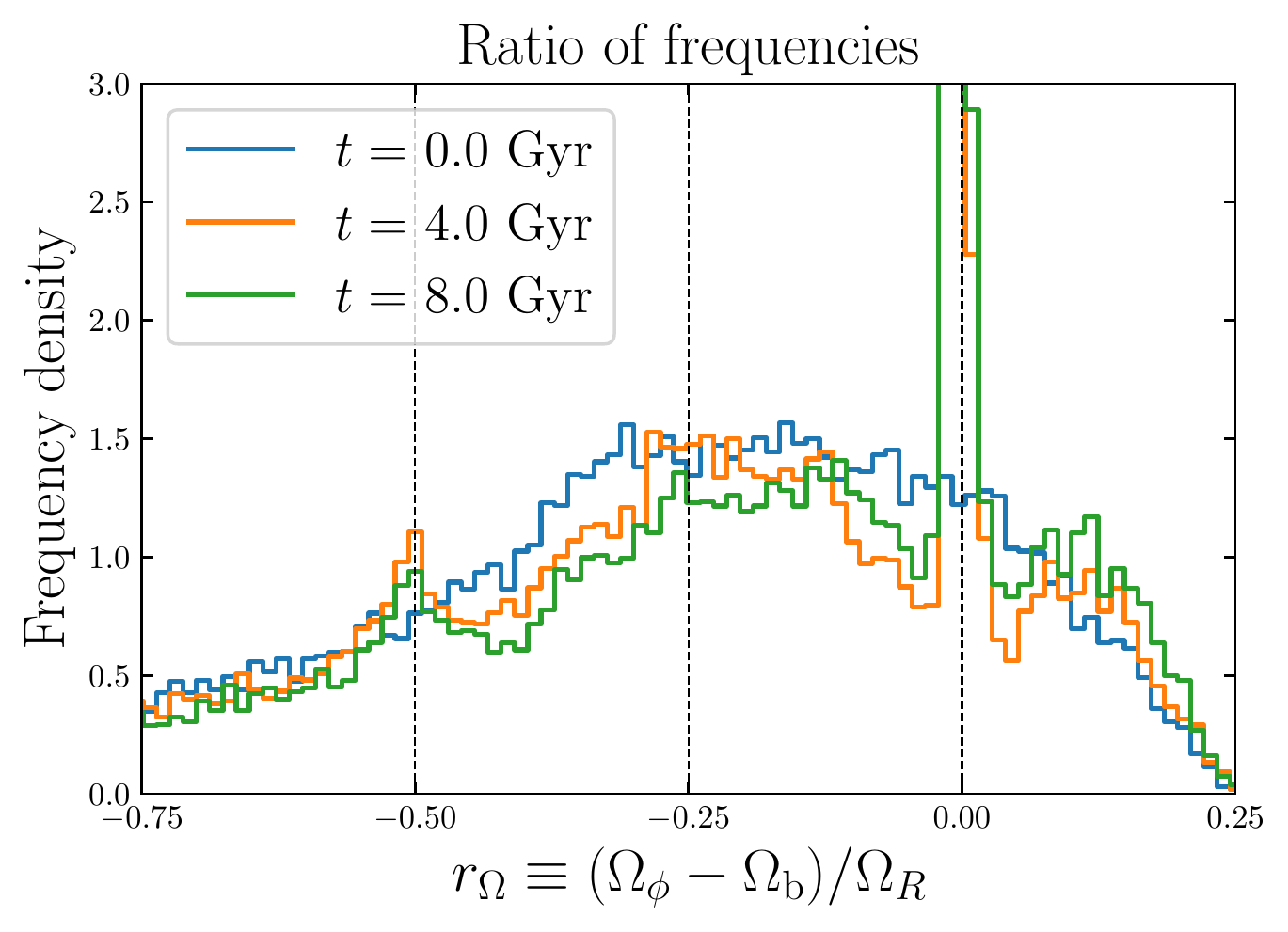}
  \caption{Distributions of $r_\Omega$, the ratio of azimuthal to radial frequencies in the frame rotating at the pattern speed $\Omega_\mathrm{b}$, at three snapshots. The distributions at $t=4$ and 8 Gyr have clear resonant peaks at $r_\Omega=0$ (corresponding to corotation) and $r_\Omega=-0.5$ (the outer Lindblad resonance). The $r_\Omega=-0.25$ or ultraharmonic resonance is also marked, although the peak is weaker here.}
   \label{fig:res_ratio_sim}
\end{figure}

\begin{figure}
  \centering
  \includegraphics[width=\columnwidth]{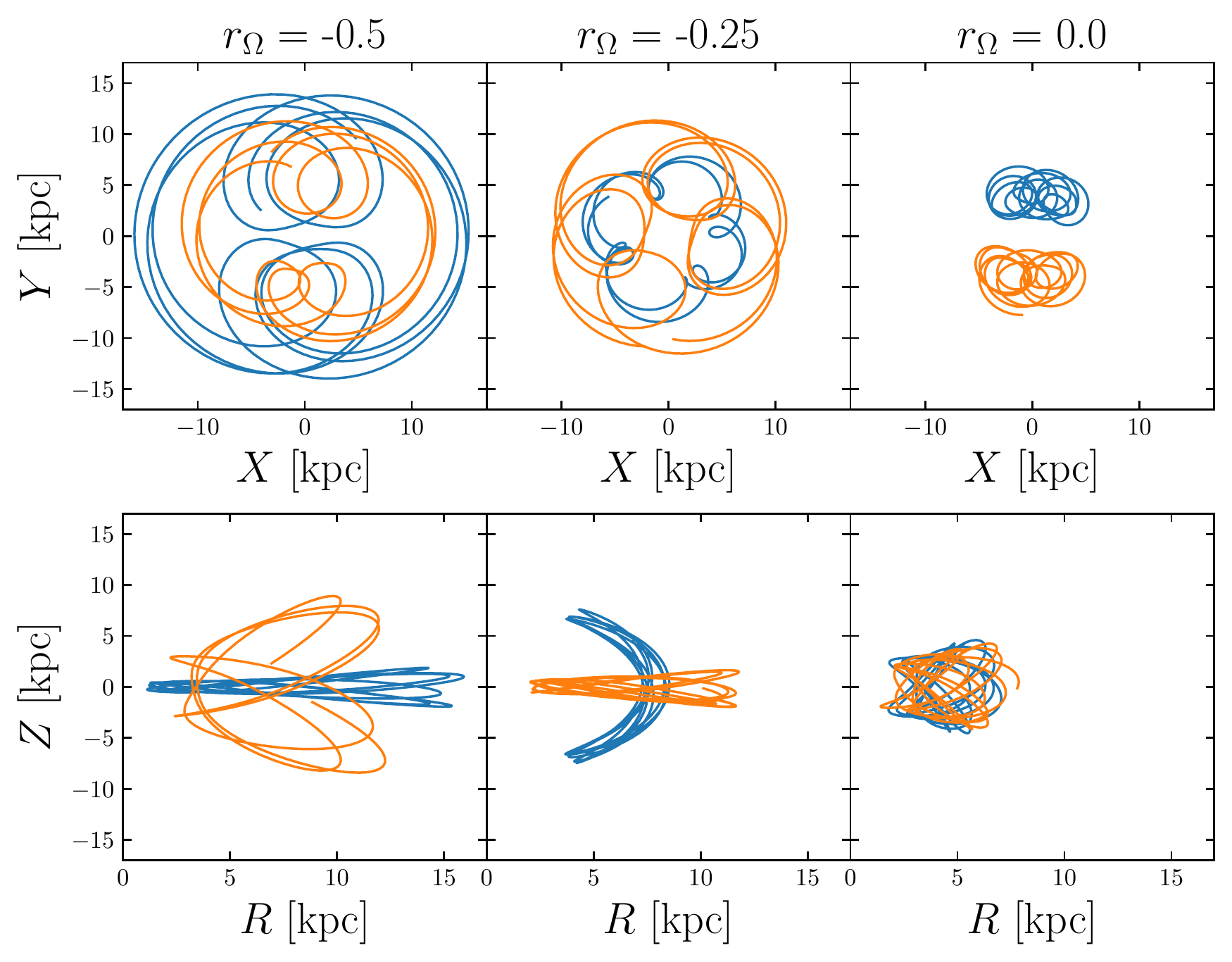}
  \caption{Orbits close to each of the three resonances marked above, plotted in the frame corotating with the bar. The Galactic disc lies in the $X-Y$ plane, with the bar's major axis along the $X$-axis. The bottom row shows the height above the disc $Z$ vs cylindrical radius $R$. The two colours show two separate orbits at each resonance. These demonstrate that trapped orbits can be both disc-like (e.g. orange, middle column) or halo-like, with large vertical excursions (e.g. orange, left-hand column).}
   \label{fig:orbits}
\end{figure}

\subsection{RVS sample weighting}
\label{section:weighting}

To correct for selection effects arising from the RVS and parallax error cuts, we apply a weighting to the data points. This is based on the ratio of our sample size to the total sample size of stars with photo-geometric distance measurements, as a function of $r$. This procedure is described in detail below.

We find the distributions in $r$ of sources in DR3 with non-zero photo-geometric distance (Sample A), and also those with $\varpi/\sigma_\varpi>10$ and non-zero RVS error (Sample B; similar to the final sample we use without more specific cuts). To undo this effect we calculate the ratio of the two distributions as a function of $r$. This ratio (Sample B/Sample A) can be seen in the top panel Fig.~\ref{fig:weighting}. Due to the RVS and parallax error cuts, this ratio has a strong peak around the Sun ($r\approx7-9$~kpc), which will likely cause spurious peaks in the energy distribution. We therefore weight our data points by the ratio Sample A/Sample B to undo the RVS and parallax error selection effects. The bottom panel of Fig.~\ref{fig:weighting} shows that this weighting largely removes the strong peak in the distribution of $r$ in our sample around the Sun, leaving a much flatter distribution.

The results can be seen in the lower two panels of Fig.~\ref{fig:E_Lz_data} and in the left-hand column of Fig.~\ref{fig:data_chevrons}. In the latter case the overdensity at $r=8$ kpc \citep[see][]{belokurov_chevrons} is almost completely removed, resulting in a much more informative distribution. For the energy-angular momentum distributions (Figs.~\ref{fig:E_Lz_data} and \ref{fig:E_comparison}) we set the weighting to zero wherever it exceeds 200 (i.e. where the function in Figs.~\ref{fig:weighting} is less than 0.005). This removes stars whose weighting is extremely large due to low sample density, and is equivalent to removing stars at $r\lesssim5$ kpc and $r\gtrsim11$ kpc.

\subsection{Energy-angular momentum space}
\label{section:E_Lz_data}

The plot of energy $E$ versus $z$-component of the angular momentum $L_z$ is a good approximation to action space ($J_r, J_\phi$). This is because energy is a proxy for radial action $J_r$, whilst $L_z$ is the azimuthal action $J_\phi$ in nearly axisymmetric potentials.

To calculate estimates of the energies $E$ of the stars in our sample, we use a modified version of the potential \texttt{MWPotential2014} described in \citet{galpy}. Similarly to \citet{belokurov_chevrons}, we set the virial mass of the Navarro-Frenk-White dark matter halo to $1\times10^{12}M_\odot$, the virial radius to 260 kpc and the concentration to 18.8, which gives a circular velocity at the Sun's radius of about 235 km\,s$^{-1}$~\citep[c.f.][]{Bl16}.

The distribution of $E$ versus $L_z$ is plotted in the top panel of Fig.~\ref{fig:E_Lz_data}. The left-handed coordinate system means that stars in the disc have $L_z>0$. The second panel shows the same distribution after unsharp masking has been applied in $E$ to reveal details of the distribution more clearly. Alternating overdensities (black) and underdensities (white) are seen most prominently at $L_z>0$ with roughly constant energy.

The weighted distributions are shown in the third and fourth panels of Fig.~\ref{fig:E_Lz_data}. The weighting has mostly removed the energy ridge at $E\approx-1.5\times10^5$ km$^2$\,s$^{-2}$, indicating that this resulted from selection effects. However, the highest energy ridge at $E\approx-1.4\times10^5$ km$^2$\,s$^{-2}$ remains, suggesting that this is a genuine feature of the distribution. It can also be seen in the dissection of the stellar halo in action space by \citet{My18}, albeit with a much smaller dataset.

\subsection{Radial phase space}
\label{section:radial_phase_space}

To isolate the region of $E$-$L_z$ space where the ridge is most prominent, we apply an angular momentum cut to include only stars with $0.25\times10^3<L_z\;\mathrm{[kpc\;km\,s^{-1}]}<1.0\times10^3$. We have checked that the results are not sensitive to the exact choice of $L_z$ cut.

We calculate the Galactocentric spherical radii and radial velocities $(r,v_r)$ for each particle.  The $(r,v_r)$ distributions are shown in Fig.~\ref{fig:data_chevrons} for stars with [Fe/H] $<-0.7$ (top row) and [Fe/H] $>-0.7$ (bottom row). While the low metallicity sample contains a mix of accreted and \textit{in situ} stars \citep{belokurov_kravtsov, conroy2022_disk, myeong2022_eccentric}, [Fe/H] $>-0.7$ is dominated by \textit{in situ} stars \citep{belokurov_chevrons}. We use the weighting described above in the left-hand column, and in the right-hand column we normalise the histograms so that each column of pixels has the same total count.

The chevron-shaped ridges reported by \citet{belokurov_chevrons} are visible, particularly for [Fe/H] < -0.7. For both metallicity cuts the most prominent feature is a roughly triangular overdensity reaching a maximum radius of $\sim7$~kpc. We find that its position and size is highly dependent on the choice of $L_z$ cut used, and it coincides with the sloping low energy overdensity at the edge of the $E$-$L_z$ distribution in Fig.~\ref{fig:E_Lz_data}. This therefore consists mostly of stars on close to circular orbits in the disc. Its shape depends on the $L_z$ cut because the energy of disc orbits is $L_z$-dependent.

More interestingly, there is a second fainter chevron that reaches $\approx10.5$ kpc, corresponding to `Chevron 1' described by \citet{belokurov_chevrons}. This is a sharp edge in the distribution, outside of which the density of stars is lower. This is visible for both metallicity cuts, albeit more faintly at [Fe/H] > -0.7 \citep[also see the bottom-left panel of Fig. 8 in][]{belokurov_chevrons}. We have found that the position of this chevron is independent of the exact $L_z$ cut used. This suggests that the orbital energies of its stars have little $L_z$ dependence (by contrast to the disc). We therefore postulate that it is a manifestation of the horizontal overdensity in Fig.~\ref{fig:E_Lz_data} at $E\approx-1.4\times10^5$ km$^2$\,s$^{-2}$. To test this we plot the phase space positions of particles with energies very close to this in the middle column of Fig.~\ref{fig:data_chevrons}. These particles form a chevron with its outer edge closely lining up with the edge of Chevron 1. This strongly suggests that the energy ridge and chevron are indeed different representations of the same entity. The fact that these stars span a wide range of radii indicates that this overdensity is not merely due to a radial selection effect around the Sun.

In summary, this is a feature at fixed energy, comprised of particles on mostly prograde orbits with a range of $L_z$. The radial phase space shows that they have apocentres of up to $\approx10$~kpc. \citet{belokurov_chevrons} suggested that this and other chevrons arose from the phase-mixing of the debris from a massive satellite which merged with the Milky Way, possibly the GSE. However, the presence of Chevron 1 at high [Fe/H] challenges this hypothesis, since very few GSE stars have such high metallicities \citep{belokurov_chevrons}.

\subsection{Configuration and frequency space}

We examine the distribution of stars in the $E$-$L_z$ overdensity as follows. We select stars in the angular momentum range $0.25<L_z$ [$10^3$ kpc km\,s$^{-1}$] $<1.0$ (between the red dashed lines in Fig.~\ref{fig:E_Lz_data}), and with energies between $E=(-1.4\pm0.02)\times10^5$ km$^2$\,s$^{-2}$. This isolates the clear ridge in the bottom panel of Fig.~\ref{fig:E_Lz_data}. We plot the cylindrical coordinates ($R,z$) of this sample in Fig.~\ref{fig:corotation_R_z}. This demonstrates the large vertical extent of the distribution of stars in this overdensity. While our sample as a whole has a large contribution from the disc at $|z|\lesssim1$ (see Fig.~\ref{fig:sample_spatial_dists}), no such excess is apparent in Fig.~\ref{fig:corotation_R_z}. This suggests that this part of the overdensity is comprised largely of stars on halo orbits. We note that more disc stars would be present if we increased the upper bound of our $L_z$ cut to include orbits with low eccentricity and inclination. However, in this study we focus on the halo-like orbits at lower $L_z$. 

We wish to consider whether the overdensity could be associated with a resonance with the bar. We therefore calculate the orbital frequencies $\Omega_i\equiv\partial H/\partial J_i$ of the stars in the axisymmetric potential. Here $H$ is the Hamiltonian and $J_i$ are the actions, calculated using the St{\"a}ckel approximation \citep[see e.g.][]{dezeeuw1985,sanders2016}. We show the distributions of the azimuthal frequency $\Omega_\phi$ in Fig.~\ref{fig:data_freqs} for stars within the $L_z$ cut described above, weighted as outlined in Section~\ref{section:weighting}. All stars within the cut are included in the black histogram, while the stars close to $E=-1.4\times10^5$ km$^2$\,s$^{-2}$ are shown in red. The stars in this overdensity have frequencies strongly peaked around 34 rad$\,$Gyr$^{-1}$, similar to some recent estimates of the pattern speed of the bar \citep[e.g.][]{binney2020}. This suggests that this overdensity may be associated with the bar's corotation resonance. We note however that the true frequencies in a rotating barred potential may differ somewhat to these estimates, so simulations with a bar are necessary for a more accurate comparison. This is what we proceed to carry out in Section~\ref{section:simulations}.

\section{Simulations}
\label{section:simulations}

We now investigate whether such a feature with the above properties can instead be produced by a rotating bar via resonant trapping. We run test particle simulations of a stellar halo-like population of particles using the galactic dynamics package \textsc{Agama} \citep{agama}. We initialise this population from a steady-state distribution function in the axisymmetric Milky Way potential described in Section~\ref{section:E_Lz_data}, before introducing a rotating bar-like perturbation which smoothly increases in strength.

\subsection{Barred potential}

To represent a bar, we use the \citet{dehnen2000} model generalised to 3 dimensions by \citet{monari2016}. This consists of a quadrupole perturbation to the potential, written in cylindrical coordinates $(R,\phi,z)$ in the form
\begin{align}
    \Phi_\mathrm{b}=A_\mathrm{b}(t)\,\mathrm{cos}(2[\phi-\phi_\mathrm{b}(t)])\,\left(\frac{R}{r}\right)^2\times\begin{cases}
      (r/R_\mathrm{b})^3-2 & r<R_\mathrm{b} \\
      -(R_\mathrm{b}/r)^3 & r\geq R_\mathrm{b},
   \end{cases}
\end{align}
where $r^2=R^2+z^2$. We increase the amplitude of the perturbation $A_\mathrm{b}(t)$ smoothly between times $t_0$ and $t_1$ according to

\begin{align}
    A_\mathrm{b}(t)&=A_f\left(\frac{3}{16}\xi^5-\frac{5}{8}\xi^3+\frac{15}{16}\xi+\frac{1}{2}\right), \\
    \xi&\equiv2\frac{t-t_0}{t_1-t_0}-1.
\end{align}
For $t<t_0$ and $t>t_1$, $A_\mathrm{b}$ equals $0$ and $A_f$ respectively. Following \citet{dehnen2000}, we parameterise the bar perturbation amplitude with the bar strength
\begin{equation}
    \alpha\equiv3\frac{A_f}{v_0^2}\left(\frac{R_\mathrm{b}}{R_0}\right)^3,
\end{equation}
where $v_0$ is the circular velocity at $R_0\equiv8$~kpc. We rotate this potential at constant pattern speed $\Omega_\mathrm{b}$. We compare simulations with pattern speeds of $\Omega_\mathrm{b}=\{35, 40, 45\}$ km\,s$^{-1}$\,kpc$^{-1}$, but focus on the $\Omega_\mathrm{b}=40$ km\,s$^{-1}$\,kpc$^{-1}$ model for most of this section. This is close to many recent estimates of the bar's pattern speed \citep[e.g.][]{wang2013,Po17,Sa19,binney2020}.

We set $\alpha=0.01$ following \citet{dehnen2000}, $t_0=2$ Gyr and $t_1=4$ Gyr, so the bar grows over a period of 2 Gyr. We run the simulation between $t=0$ and $t_f=8$ Gyr. This is roughly the duration of the period of the Milky Way's history since the GSE merger, over which no major satellite has been accreted \citep[e.g.][]{belokurov2018,kruijssen2020}. We set the bar radius to $R_\mathrm{b}=2$ kpc. While this is likely smaller than the Milky Way's bar at present \citep[e.g.][]{hammersley1994,wegg2015,lucey2022}, it may be more realistic for the period over which it was formed \citep{rosas-guevara2022} .

\subsection{Distribution function}

We generate a halo-like population of stars using the double power law distribution function implemented in \textsc{Agama} by \citet{agama}, which is a generalisation of ones introduced by \citet{posti2015} and \citet{Wi15}:
\begin{align}
    f(\textbf{\textit{J}})&=\frac{M}{(2\pi J_0)^3}\left[1+\left(\frac{J_0}{h(\textbf{\textit{J}})}\right)^\eta\right]^{\Gamma/\eta}\left[1+\left(\frac{h(\textbf{\textit{J}})}{J_0}\right)^\eta\right]^{-B/\eta},\\
    h(\textbf{\textit{J}})&\equiv J_r+|J_\phi|+J_z.
\end{align}
Here, $J_0$ is the characteristic total action $|\textbf{\textit{J}}|$ of orbits near the break radius of the double-power law profile \citep{posti2015}.

We wish to choose values of $J_0$ and the power law indices $\Gamma$ and $B$ to roughly reproduce the density profile of the Milky Way's stellar halo. Multiple studies have found that the density profile can be fitted by a double power law with inner slope $\gamma\sim2.5$ and outer slope $\beta\sim4.5$, with a break radius at $r\approx25$~kpc \citep[e.g.][]{Wa09,deason2011_halo,faccioli2014_halo,pila-diez2015}. We find that choosing $J_0=3500$ kpc~km\,s$^{-1}$ and $(\Gamma,B)=(2.4,4.6)$ roughly recovers the values for the Milky Way. We set the steepness of the break to $\eta=10$, which results in the density power law slope changing between radii of $\sim10-40$~kpc.

\subsection{Results}

We integrate the orbits of these stars in the combined potential of the Milky Way and the growing rotating bar for time $t_f$. We calculate the energy $E$ and angular momentum $L_z$ of the star particles in each snapshot, and show these distributions in Fig.~\ref{fig:E_Lz_sim}. At the beginning of the simulation, there is a large concentration of stars at small $L_z$ but otherwise no structure. However, at $t=4$ and 8 Gyr horizontal striations are visible at $L_z>0$, corresponding to specific energies where there are overdensities and underdensities of stars. These stripes resemble those seen in the data in Fig.~\ref{fig:E_Lz_data}, as they occupy a similar region of $E$-$L_z$ space. Since the initial distribution function is symmetric in $L_z$, the fact that these simulated ridges occur only at $L_z>0$ implies that they must be related to the rotation of the bar.

The third panel of Fig.~\ref{fig:E_Lz_sim} shows how the star particles move through $E$-$L_z$ space. The red streaks mark the paths taken by a selection of stars, where the motion is towards the top-right. This results in a depletion of stars at $L_z\leq0$ (white pixels) and an enhancement in their density in the ridges at $L_z>0$. This behaviour strongly resembles that expected for a steady but rotating non-axisymmetric potential, where stars are constrained to move along lines of gradient $\Omega_\mathrm{b}$ in the $E$ vs $L_z$ plane due to conservation of the Jacobi integral. We see similar behaviour here, though we note that the bar perturbation grows in magnitude between $t=2$ and 4 Gyr, so the Jacobi integral is not exactly conserved during this period. We have checked that the gradient of these streaks is indeed approximately $\Omega_\mathrm{b}$.

We calculate the instantaneous radial and azimuthal frequencies of the star particles as follows. We integrate their orbits from their phase space positions in each snapshot, using the corresponding potential with constant pattern speed. The frequencies are calculated from the radial and azimuthal motion averaged over several orbital periods. In the bottom panel of Fig.~\ref{fig:E_Lz_sim} we plot only particles with azimuthal frequencies satisfying $|\Omega_\phi-\Omega_\mathrm{b}|/\Omega_\mathrm{b}<0.001$ (i.e. stars very close to the corotation resonance). The distribution of these stars exactly coincides with the lowest energy overdensity seen in the panel above, at $E\approx-1.5\times10^5$ km$^2$\,s$^{-2}$. This overdensity can therefore be associated with the corotation resonance.

We show the radial phase space of the simulations in Fig.~\ref{fig:sim_chevrons}, where we have applied the same radius and angular momentum cuts as on the data in Fig.~\ref{fig:data_chevrons} ($4<r\;\mathrm{[kpc]}<12$, $0.25\times10^3<L_z\;\mathrm{[kpc\;km\,s^{-1}]}<1.0\times10^3$). The top and bottom rows show the start and end of the simulation, at $t=0$ and 8 Gyr respectively. The three columns use different normalisations.

The initial distribution function has a smooth radial phase space, with no substructure visible. However, by the end of the simulation a wedge-shaped overdense region emerges, with its tip at $r\approx10$ kpc. This bears a strong resemblance to Chevron 1 visible in the data in Fig.~\ref{fig:data_chevrons}

To assess how the change in $L_z$ depends on the amplitude of vertical motion of the stars, we calculate the vertical actions $J_z$ using the St{\"a}ckel Fudge approximation \citep{binney2012} in an axisymmetrised potential (without the bar). The distributions of $J_z$ vs $L_z$ are shown in Fig.~\ref{fig:Jz_Lz} at the beginning and end of the simulation. The bottom panel shows the ratio of the two distributions (i.e. the difference in log density). As expected from Fig.~\ref{fig:E_Lz_sim}, at the end of the simulation the average $L_z$ is positive, as stars move away from around $L_z=0$. Fig.~\ref{fig:Jz_Lz} shows that this has $J_z$ dependence, with particles at low $J_z$ experiencing greater increases in $L_z$ on average. This is likely to be in part due to the fact that stars at low $J_z$ tend to have lower energy and hence interact more strongly with the bar.

Fig.~\ref{fig:Jz_Lz} demonstrates that the bar is capable of creating a population of stars with low $J_z$ and positive mean $L_z$ from an initially non-rotating distribution. We note that this bears some similarity to the population ultra metal-poor stars discussed by \citet{sestito2019}. A significant fraction of these stars are on prograde orbits with $J_z\lesssim100$ kpc\,km\,s$^{-1}$, comparable to the distribution in the middle panel of Fig.~\ref{fig:Jz_Lz}. Our results suggest that the Milky Way's bar could be responsible for creating the observed distribution of these stars and thus alleviate the need to invoke an early emergence of a prehistoric metal-poor disc \citep[see e.g.][]{sestito2019,DiMatteo2020,Mardini2022}.


We compute the spherically averaged median azimuthal velocity $v_\phi$ as a function of cylindrical radius $R$. This is plotted in Fig.~\ref{fig:density_profile_change} at three snapshots. The distribution changes from having no net spin to having a positive median $v_\phi$, with larger values at small radii. This is expected from Figs.~\ref{fig:E_Lz_sim} and \ref{fig:Jz_Lz}, which show that by $t=8$ Gyr there is an excess of star particles at $L_z>0$, particularly at low energies.

In Fig.~\ref{fig:density_profile_change} we also plot data from data release 17 (DR17) of the APOGEE survey \citep{apogee,apogee_dr17} for comparison. We follow closely the steps outlined in \citet{belokurov_kravtsov} to select unproblematic red giants stars with small chemical and kinematic uncertainties. In addition, we apply [Al/Fe]~$<0.1$ and [Fe/H]~$<-1.2$ cuts to limit our sample to predominantly accreted halo stars. These APOGEE low-metallicity data points qualitatively follow a similar trend to the simulation, with a decreasing median $v_\phi$ with increasing $R$.


\begin{figure}
  \centering
  \includegraphics[width=\columnwidth]{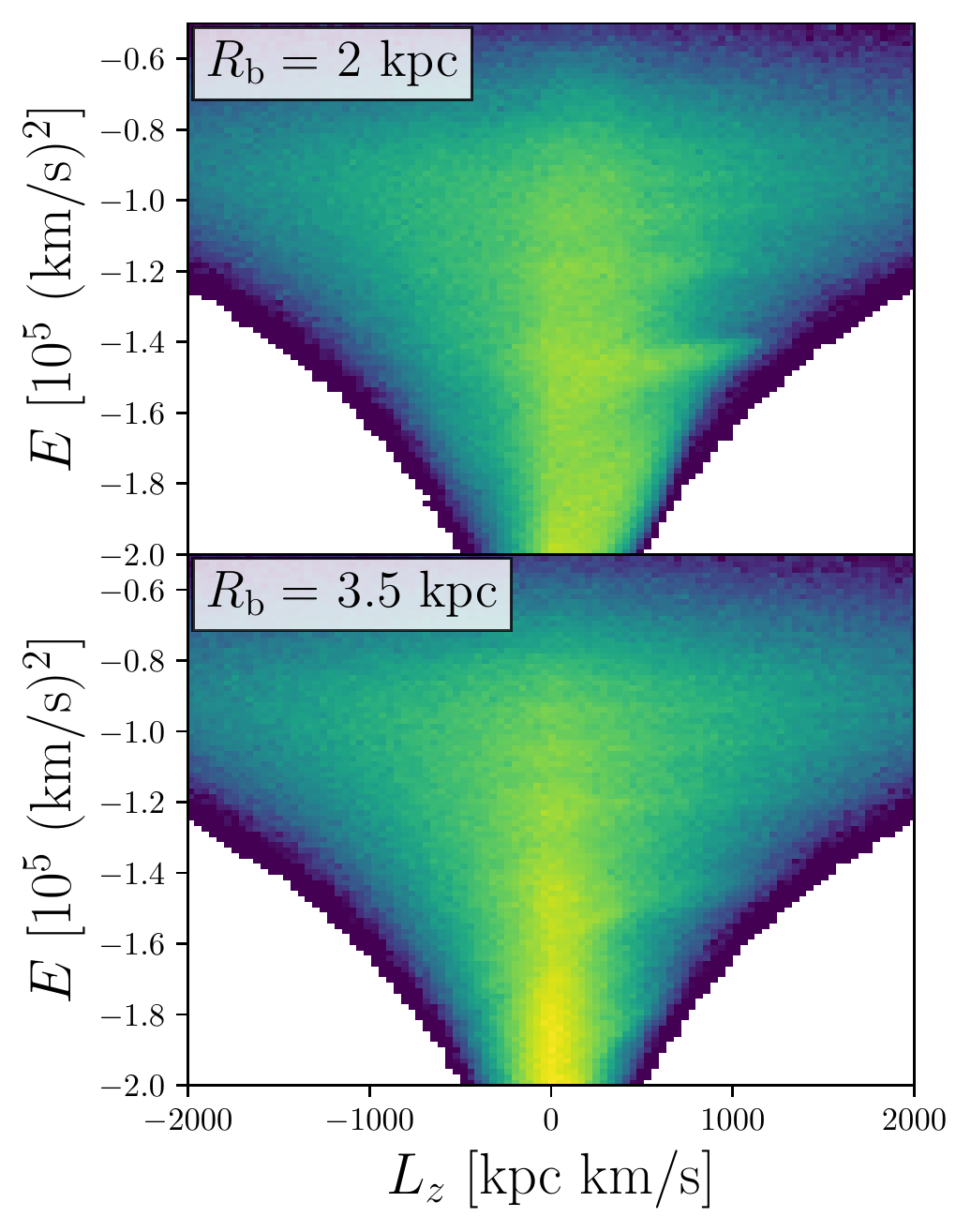}
  \caption{Energy-angular momentum space at the end of the simulation with a bar length of $R_\mathrm{b}=2$~kpc (top panel) and 3~kpc (bottom panel). The overdensities are still clearly visible with the longer bar, albeit with somewhat lower contrast.}
   \label{fig:bar_length_comparison}
\end{figure}

\begin{figure*}
  \centering
  \includegraphics[width=\textwidth]{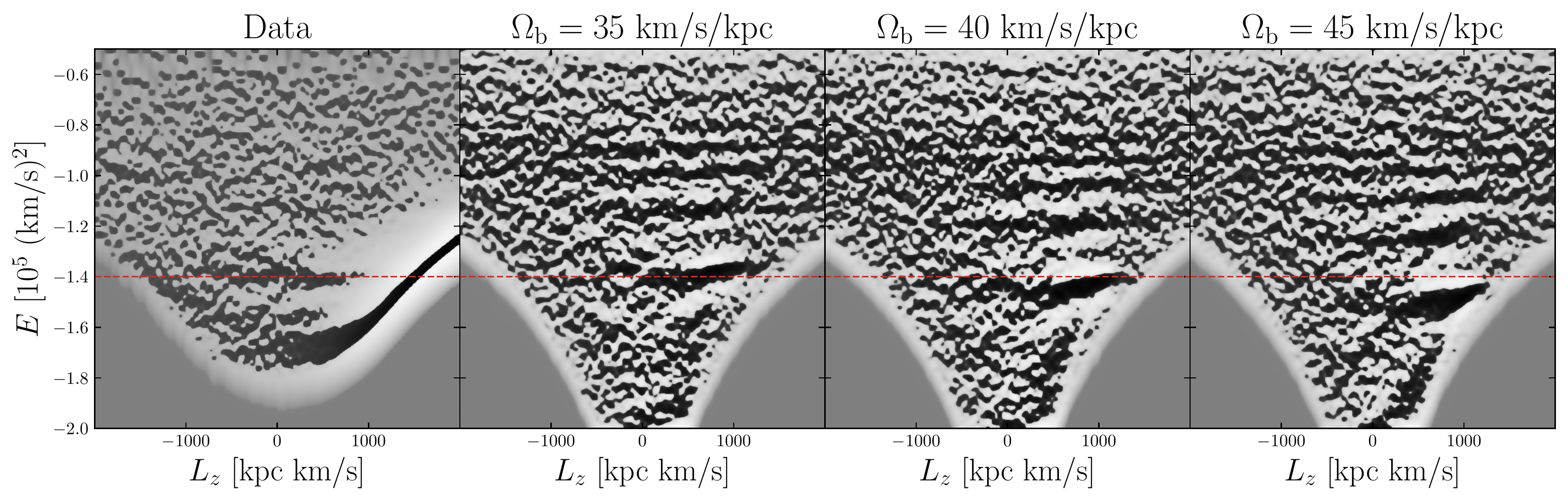}
  \caption{$E$-$L_z$ distributions of the data (left-hand panel) and simulations with three different pattern speeds (right three panels). As in the bottom panel of Fig.~\ref{fig:E_Lz_data}, the data have been weighted and unsharp filtering has been applied to each distribution in energy. Black (white) pixels denote overdensities (underdensities). The red dashed line marks the approximate energy of the horizontal ridge in the data at $L_z>0$. As the pattern speed is increased, the horizontal ridges in the simulations decrease in energy (and increase in frequency). The most prominent one (due to the corotation resonance) roughly aligns with the ridge in the data at a pattern speed of $\Omega_\mathrm{b}\approx35-40$ km\,s$^{-1}$\,kpc$^{-1}$.}
   \label{fig:E_comparison}
\end{figure*}

In Fig.~\ref{fig:Omega_sim}, we plot the final versus initial values of $\Omega_\phi$ and $\Omega_R$ for the simulated star particles with the $L_z$ and $r$ cuts used for Fig.~\ref{fig:sim_chevrons}. While many particles' frequencies do not change between the start and end of the simulation, narrow horizontal stripes in the panels of Fig.~\ref{fig:Omega_sim} indicate that the final frequency distributions are tightly clustered around particular values. This includes $\Omega_\phi=40$~km\,s$^{-1}$\,kpc$^{-1}$, which is equal to the pattern speed of the bar. Hence these particles are trapped around the corotation resonance. The red points show particles with energies close to that of the lowest energy overdensity in Fig.~\ref{fig:E_Lz_sim}. Many of these are located at or very close to the corotation resonance, as expected from Fig.~\ref{fig:E_Lz_sim}. To show this and other resonances more clearly we compute the ratio of azimuthal and radial frequencies in the frame corotating with the bar, $r_\Omega\equiv(\Omega_\phi-\Omega_\mathrm{b})/\Omega_R$. The distributions of this quantity are plotted in Fig.~\ref{fig:res_ratio_sim} at three different times.

This shows that the frequencies are clustered around multiple resonances with the bar. While the initial distribution is relatively smooth, the intermediate and final distributions have multiple sharp peaks located at particular values of $r_\Omega$. This includes the corotation resonance $r_\Omega=0$ and the outer Lindblad resonance $r_\Omega=-0.5$, with the corotation peak begin particularly strong. The distributions at 4 and 8 Gyr are very similar, showing that the trapping takes place during the growth phase of the bar ($2<t/\mathrm{Gyr}<4$) and that the resonant peaks are largely unchanged while the bar is steadily rotating with constant strength.

We show a selection of orbits near three of the resonances in Fig.~\ref{fig:orbits}. The top row shows the orbits in the plane of the disc in the frame corotating with the bar (whose major axis lies along the $X$-axis). Stars near the corotation resonance (right-hand column) do not cross the bar's major axis, instead orbiting on one side only. The bottom row shows the vertical vs radial behaviour of the orbits. While some resonant orbits remain close to the disc ($Z=0$ plane), others have large vertical excursions, reaching over 5 kpc above and below the disc. Hence we see that stars on halo-like orbits can contribute to the resonant peaks seen in Fig.~\ref{fig:res_ratio_sim}. Note that the blue orbit in the middle column also has a resonance between its $R$ and $Z$ motion.

\subsubsection{Dependence on bar length}

To check how our results depend on the bar length $R_\mathrm{b}$, we run another simulation with $R_\mathrm{b}=3.5$~kpc. This is the value used by \citet{monari2016}, and is a good fit to the current (rather than initial) size of the Milky Way's bar. We maintain a bar strength of $\alpha=0.01$, so the strength of the perturbation near the Sun is unchanged. We show a comparison of the $E$-$L_z$ space of the two simulations in Fig.~\ref{fig:bar_length_comparison}.

We see that the ridge overdensities with roughly constant energy at $L_z>0$ remain visible with the longer bar. They are slightly less prominent than with $R_\mathrm{b}=2$~kpc, with fewer stars being moved away from $L_z\approx0$. However, the energies of the overdensities are unaffected by the change in bar size, so our analysis is independent of the exact choice of bar model. We also note that the strength of the overdensities can be changed by a time-dependent pattern speed through the process of resonance sweeping \citep[see e.g.][]{chiba2021}. We plan to address this in a future work.

\subsection{Comparison with data}
\label{section:comparison}

We wish to estimate the approximate pattern speed of the bar which can best reproduce the observed structures in Fig.~\ref{fig:E_Lz_data}. For this purpose, we run simulations with constant pattern speeds $\Omega_\mathrm{b}=\{35, 40, 45\}$ km\,s$^{-1}$\,kpc$^{-1}$ with the same background potential and initial distribution function. We calculate the energy $E$ and angular momentum $L_z$ of the particles, and apply unsharp masking to the $E$-$L_z$ distributions similar to that used in Fig.~\ref{fig:E_Lz_data}. The filtered $E$-$L_z$ distributions for the data and three simulations are shown in Fig.~\ref{fig:E_comparison} with the approximate energies of the observed ridges marked by red dashed lines.

As the pattern speed $\Omega_\mathrm{b}$ is increased, the frequencies of the resonances increase. This results in a decrease of the energies of the corresponding overdensities in $E$-$L_z$ space. Of the three simulations, the closest match is at a pattern speed of $\Omega_\mathrm{b}=35$ km\,s$^{-1}$\,kpc$^{-1}$, at which the energy of the corotation ridge is $-1.4\times10^5$~km$^2$\,s$^{-2}$. The $\Omega_\mathrm{b}=40$ km\,s$^{-1}$\,kpc$^{-1}$ model is also a reasonably good fit, while at 45 km\,s$^{-1}$\,kpc$^{-1}$ the energy of the corotation ridge is too low to match the data.

We therefore conclude that pattern speeds in the range $35-40$ km\,s$^{-1}$\,kpc$^{-1}$ produce good matches between the energies of the observed and simulated ridges in $E$-$L_z$ space. This is in agreement with most recent studies that constrain the pattern speed. For example, \citet{chiba2021} and \citet{binney2020} found that $\Omega_\mathrm{b}\approx35.5$ and $36$ km\,s$^{-1}$\,kpc$^{-1}$ respectively were favoured, while \citet{Po17}, \citet{wang2013} and \citet{Sa19} preferred values in the range $39-41$ km\,s$^{-1}$\,kpc$^{-1}$.

\section{Conclusions}
\label{section:conclusions}

Bars are important drivers of morphological changes in galaxies. The Milky Way is no exception. Its massive bar extends to around Galactocentric distances of $\approx 5$ kpc and contains $\approx 30-40$ per cent of the total stellar mass in the Galaxy. This hefty structure has long been known to drive evolution in the disc through its resonances. These are the locations where a combination of the stars' natural frequencies is equal to the forcing frequency or pattern speed of the bar. At the resonances, secular changes in the orbits occur, causing long term effects in the stellar populations.

Data from the {\it Gaia} satellite has reinvigorated the study of substructure in the disc. Many of the notches, ridges and ripples in phase space have been interpreted as the imprints of resonances with the bar~\citep[e.g.,][]{khoperskov2020, trick2021,Tr22}, though the effects of spirality driven by the bar are also possible~\citep{Hu18}. The Hercules and Sirius streams in the solar neighbourhood were previously associated with the outer Lindblad resonance of the bar~\citep{Ka91, De98}. This requires the bar to have a high pattern speed. Nowadays, the corotation resonance of a slow bar with pattern speed $\Omegab \approx 40$ km\,s$^{-1}$\,kpc$^{-1}$ is seen as the more likely explanation~\citep[e.g.][]{Mo19}. There have also been hints that the bar may affect both streams~\citep{Ha16} and stellar populations in the halo~\citep[e.g.,][]{Mor15,My18,Sc19}. 

In this paper, we have identified a distinctive feature in the phase space distribution of halo stars. It is a prominent ridge at energies $E\approx-1.4\times10^5$ km$^2$\,s$^{-2}$ and with $L_z>0$. This is apparent when data from the \textit{Gaia} Data Release 3 (DR3) Radial Velocity Spectrometer (RVS) sample is plotted in energy versus angular momentum ($E$-$L_z$) space. It was also seen by \citet{My18}, albeit with a much smaller dataset. Further ridges are also present before weighting, but they are likely artefacts caused by the selection function. The prominent ridge corresponds to a chevron-shaped overdensity in radial velocity versus radius ($v_r$-$r$) space previously reported by \citet{belokurov_chevrons}. it resembles those produced by phase mixing the debris of a merged satellite galaxy \citep[e.g.][]{filmore1984,sanderson2013,dong-paez2022,belokurov_chevrons}. However, the structure persists at both high and low metallicity ([Fe/H] $>-0.7$ and $<-0.7$), suggesting that the chevron is not composed of stars accreted from \textit{Gaia} Sausage-Enceladus or elsewhere. It is natural to seek a dynamical solution for its provenance, as stars of all metallicities are affected.
    
To understand its origin, we run test particle simulations of a stellar halo population of particles under the influence of a rotating bar. After generating a steady-state distribution in an axisymmetric potential, we smoothly increase the bar strength over a period of 2 Gyr. The bar remains steady for the final 4 Gyrs of its life. The resultant distribution of stars in $E$-$L_z$ and $v_r$-$r$ space shows features similar to the data. Particles move from around $L_z\approx0$ into ridges at $L_z>0$, with both $E$ and $L_z$ increasing, so as to approximately conserve the Jacobi integral. When plotted in radial phase space, the ridges appear as chevrons. At the end of the simulation, the azimuthal and radial frequencies $\Omega_\phi$ and $\Omega_R$ are strongly clustered about certain values. These correspond to resonances with the bar, especially the corotation resonance and outer Lindblad resonance. It is the corotation resonance that is responsible for the most prominent ridge in $E$-$L_z$ space. We plot the orbits of particles close to these resonances. Some are disc-like, but some are halo-like with large vertical excursions from the Galactic plane.

The bar spins up the inner stellar halo. We show this by calculating the spherically averaged median azimuthal velocity $v_\phi$ as a function of cylindrical radius $R$ in our simulations. The average $v_\phi$ at $R\lesssim10$ kpc is increased, so the simulated stellar halo gains a net spin. The observational data on the metal-poor star subsample from APOGEE shows qualitatively the same trend as the simulations, with the median $v_\phi$ decreasing from $\approx70$ to 10 km\,s$^{-1}$ between Galactocentric radii $R$ of 3 and 9 kpc. The effect of a bar spinning up a spheroidal distribution has been extensively studied in relation to the dark halo \citep[e.g.][]{weinberg1985,athanassoula2002} and the bulge \citep[e.g.][]{saha2012,saha2016}. Dynamical friction acts on the bar, transferring its angular momentum to the halo and reducing the pattern speed \citep{hernquist1992,debattista2000}. However, this process is not simple; \citet{chiba2022} showed that the dynamical friction can oscillate, resulting in the pattern speed oscillating while it decays. The efficiency of the angular momentum transfer depends on the mass of the spheroidal distribution, with less massive bulges being spun up more easily \citep{saha2016}. We therefore note that while our test particle simulation provides an illustration of this effect, an $N$-body simulation with a live bar, disc and halo is required to quantify it and fully capture its complexity.

The distribution of the angular momentum becomes asymmetric. This asymmetry is more pronounced for stars with low $J_z$. As a result, a population of stars on seemingly disc-like orbits is forged from the initially non-rotating and smooth halo distribution. These orbits are similar to those previously reported for metal-poor, very metal-poor and extremely metal-poor stars in studies advocating a disc-like state for the high-redshift Galaxy \citep[see e.g.][]{sestito2019,DiMatteo2020,Mardini2022}. If following our results, the necessity for a pre-historic disc is removed, a different picture emerges. Before spinning up into a cogerently rotating disc at metallicities $-1.5<$[Fe/H]$<-1$, the early Milky Way was instead a kinematically hot and messy place in line with the analysis by \citet{belokurov_kravtsov}. This pre-disc Milky Way population, dubbed {\it Aurora}, is strongly centrally concentrated and thus lies predominantly within Solar radius and has a modest net spin \citep[see][]{belokurov_kravtsov}. Similar conclusions as to the state of the high-redshift Galaxy are reached in recent observational \citep[][]{conroy2022_disk,Rix2022,myeong2022_eccentric} and theoretical \citep[][]{Gurvich2023,Hopkins2023} studies.

We repeat our simulation with different bar pattern speeds in the range 35-45 km\,s$^{-1}$\,kpc$^{-1}$. As the pattern speed of the bar is increased, the frequencies of the resonances increase. This decreases the energy of the horizontal ridges in $E$ vs $L_z$ space, as well as the radial extent of the corresponding chevron over-densities in $v_r$ vs $r$ space. We find that, with our choice of potential, a pattern speed of $\Omega_\mathrm{b}\approx35-40$ km\,s$^{-1}$\,kpc$^{-1}$ is required to match the energy of the most prominent ridge in $E$-$L_z$ space. This is consistent with most recent estimates of the Milky Way bar's pattern speed, most of which are $35-41$ km\,s$^{-1}$\,kpc$^{-1}$ \citep[e.g.][]{wang2013,Po17,Sa19,binney2020,chiba2021_treering}. While our results suggest that 35 km\,s$^{-1}$\,kpc$^{-1}$ may be slightly favoured over 40 km\,s$^{-1}$\,kpc$^{-1}$, a more detailed study is required to include consideration of uncertainties in the potential.

There are a number of avenues for further exploration. First, this offers up a new tool to study the Galactic bar. Its formation epoch and its evolution will imprint themselves on the stellar halo as well as the disc. If the pattern speed of the bar is slowing because of dynamical friction exerted by the dark halo, this may induce detectable features in the ridges of the stellar halo populations.  Secondly, the Lindblad and other resonances may also be the sites of depletions and enhancements in the action space of the stellar halo, though these have not yet been unambiguously associate with structures in the data. Thirdly, the bar may also create resonant features in the dark halo. This may be important if the substructures coincide with the solar neighbourhood, as this may enhance the signal expected in direct detection experiments in Earth-borne laboratories. We are actively pursuing these investigations and will report results in a future work.

\section*{Acknowledgements}
We thank the anonymous referee for very helpful comments that have improved this manuscript. We are grateful to the Cambridge Streams group and Hans-Walter Rix for insightful discussions during this study. AMD and EYD thank the Science and Technology Facilities Council (STFC) for PhD studentships.

This work has made use of data from the European Space Agency (ESA) mission
{\it Gaia} (\url{https://www.cosmos.esa.int/gaia}), processed by the {\it Gaia} Data Processing and Analysis Consortium (DPAC,
\url{https://www.cosmos.esa.int/web/gaia/dpac/consortium}). Funding for the DPAC has been provided by national institutions, in particular the institutions participating in the {\it Gaia} Multilateral Agreement.

This research made use of Astropy,\footnote{http://www.astropy.org} a community-developed core Python package for Astronomy \citep{astropy:2013, astropy:2018}. This work was funded by UKRI grant 2604986. For the purpose of open access, the author has applied a Creative Commons Attribution (CC BY) licence to any Author Accepted Manuscript version arising.

\section*{Data Availability}

This study uses publicly available \textit{Gaia} data.



\bibliographystyle{mnras}
\bibliography{refs} 

\begin{thebibliography}{}
\makeatletter
\relax
\def\mn@urlcharsother{\let\do\@makeother \do\$\do\&\do\#\do\^\do\_\do\%\do\~}
\def\mn@doi{\begingroup\mn@urlcharsother \@ifnextchar [ {\mn@doi@}
  {\mn@doi@[]}}
\def\mn@doi@[#1]#2{\def\@tempa{#1}\ifx\@tempa\@empty \href
  {http://dx.doi.org/#2} {doi:#2}\else \href {http://dx.doi.org/#2} {#1}\fi
  \endgroup}
\def\mn@eprint#1#2{\mn@eprint@#1:#2::\@nil}
\def\mn@eprint@arXiv#1{\href {http://arxiv.org/abs/#1} {{\tt arXiv:#1}}}
\def\mn@eprint@dblp#1{\href {http://dblp.uni-trier.de/rec/bibtex/#1.xml}
  {dblp:#1}}
\def\mn@eprint@#1:#2:#3:#4\@nil{\def\@tempa {#1}\def\@tempb {#2}\def\@tempc
  {#3}\ifx \@tempc \@empty \let \@tempc \@tempb \let \@tempb \@tempa \fi \ifx
  \@tempb \@empty \def\@tempb {arXiv}\fi \@ifundefined
  {mn@eprint@\@tempb}{\@tempb:\@tempc}{\expandafter \expandafter \csname
  mn@eprint@\@tempb\endcsname \expandafter{\@tempc}}}

\bibitem[\protect\citeauthoryear{{Abdurro'uf} et~al.,}{{Abdurro'uf}
  et~al.}{2022}]{apogee_dr17}
{Abdurro'uf} et~al., 2022, \mn@doi [\apjs] {10.3847/1538-4365/ac4414}, \href
  {https://ui.adsabs.harvard.edu/abs/2022ApJS..259...35A} {259, 35}

\bibitem[\protect\citeauthoryear{{Arnold}}{{Arnold}}{1978}]{Ar78}
{Arnold} V.~I.,  1978, {Mathematical methods of classical mechanics}

\bibitem[\protect\citeauthoryear{{Astropy Collaboration} et~al.,}{{Astropy
  Collaboration} et~al.}{2013}]{astropy:2013}
{Astropy Collaboration} et~al., 2013, \mn@doi [\aap]
  {10.1051/0004-6361/201322068}, \href
  {http://adsabs.harvard.edu/abs/2013A%26A...558A..33A} {558, A33}

\bibitem[\protect\citeauthoryear{{Astropy Collaboration} et~al.,}{{Astropy
  Collaboration} et~al.}{2018}]{astropy:2018}
{Astropy Collaboration} et~al., 2018, \mn@doi [\aj] {10.3847/1538-3881/aabc4f},
  \href {https://ui.adsabs.harvard.edu/abs/2018AJ....156..123A} {156, 123}

\bibitem[\protect\citeauthoryear{{Athanassoula}}{{Athanassoula}}{2002}]{athanassoula2002}
{Athanassoula} E.,  2002, \mn@doi [\apjl] {10.1086/340784}, \href
  {https://ui.adsabs.harvard.edu/abs/2002ApJ...569L..83A} {569, L83}

\bibitem[\protect\citeauthoryear{{Bailer-Jones}, {Rybizki}, {Fouesneau},
  {Demleitner}  \& {Andrae}}{{Bailer-Jones} et~al.}{2021}]{bailer-jones2021}
{Bailer-Jones} C.~A.~L.,  {Rybizki} J.,  {Fouesneau} M.,  {Demleitner} M.,
  {Andrae} R.,  2021, \mn@doi [\aj] {10.3847/1538-3881/abd806}, \href
  {https://ui.adsabs.harvard.edu/abs/2021AJ....161..147B} {161, 147}

\bibitem[\protect\citeauthoryear{{Belokurov} \& {Kravtsov}}{{Belokurov} \&
  {Kravtsov}}{2022}]{belokurov_kravtsov}
{Belokurov} V.,  {Kravtsov} A.,  2022, \mn@doi [\mnras]
  {10.1093/mnras/stac1267}, \href
  {https://ui.adsabs.harvard.edu/abs/2022MNRAS.514..689B} {514, 689}

\bibitem[\protect\citeauthoryear{{Belokurov}, {Erkal}, {Evans}, {Koposov}  \&
  {Deason}}{{Belokurov} et~al.}{2018}]{belokurov2018}
{Belokurov} V.,  {Erkal} D.,  {Evans} N.~W.,  {Koposov} S.~E.,   {Deason}
  A.~J.,  2018, \mn@doi [\mnras] {10.1093/mnras/sty982}, \href
  {https://ui.adsabs.harvard.edu/abs/2018MNRAS.478..611B} {478, 611}

\bibitem[\protect\citeauthoryear{{Belokurov}, {Vasiliev}, {Deason}, {Koposov},
  {Fattahi}, {Dillamore}, {Davies}  \& {Grand}}{{Belokurov}
  et~al.}{2023}]{belokurov_chevrons}
{Belokurov} V.,  {Vasiliev} E.,  {Deason} A.~J.,  {Koposov} S.~E.,  {Fattahi}
  A.,  {Dillamore} A.~M.,  {Davies} E.~Y.,   {Grand} R. J.~J.,  2023, \mn@doi
  [\mnras] {10.1093/mnras/stac3436}, \href
  {https://ui.adsabs.harvard.edu/abs/2023MNRAS.518.6200B} {518, 6200}

\bibitem[\protect\citeauthoryear{{Benjamin} et~al.,}{{Benjamin}
  et~al.}{2005}]{benjamin2005}
{Benjamin} R.~A.,  et~al., 2005, \mn@doi [\apjl] {10.1086/491785}, \href
  {https://ui.adsabs.harvard.edu/abs/2005ApJ...630L.149B} {630, L149}

\bibitem[\protect\citeauthoryear{{Binney}}{{Binney}}{2012}]{binney2012}
{Binney} J.,  2012, \mn@doi [\mnras] {10.1111/j.1365-2966.2012.21757.x}, \href
  {https://ui.adsabs.harvard.edu/abs/2012MNRAS.426.1324B} {426, 1324}

\bibitem[\protect\citeauthoryear{{Binney}}{{Binney}}{2020}]{binney2020}
{Binney} J.,  2020, \mn@doi [\mnras] {10.1093/mnras/staa1103}, \href
  {https://ui.adsabs.harvard.edu/abs/2020MNRAS.495..895B} {495, 895}

\bibitem[\protect\citeauthoryear{{Binney} \& {Tremaine}}{{Binney} \&
  {Tremaine}}{2008}]{binney_tremaine}
{Binney} J.,  {Tremaine} S.,  2008, {Galactic Dynamics: Second Edition}

\bibitem[\protect\citeauthoryear{{Binney}, {Gerhard}, {Stark}, {Bally}  \&
  {Uchida}}{{Binney} et~al.}{1991}]{Bi91}
{Binney} J.,  {Gerhard} O.~E.,  {Stark} A.~A.,  {Bally} J.,   {Uchida} K.~I.,
  1991, \mn@doi [\mnras] {10.1093/mnras/252.2.210}, \href
  {https://ui.adsabs.harvard.edu/abs/1991MNRAS.252..210B} {252, 210}

\bibitem[\protect\citeauthoryear{{Bissantz}, {Englmaier}  \&
  {Gerhard}}{{Bissantz} et~al.}{2003}]{Bi03}
{Bissantz} N.,  {Englmaier} P.,   {Gerhard} O.,  2003, \mn@doi [\mnras]
  {10.1046/j.1365-8711.2003.06358.x}, \href
  {https://ui.adsabs.harvard.edu/abs/2003MNRAS.340..949B} {340, 949}

\bibitem[\protect\citeauthoryear{{Bland-Hawthorn} \&
  {Gerhard}}{{Bland-Hawthorn} \& {Gerhard}}{2016}]{Bl16}
{Bland-Hawthorn} J.,  {Gerhard} O.,  2016, \mn@doi [\araa]
  {10.1146/annurev-astro-081915-023441}, \href
  {https://ui.adsabs.harvard.edu/abs/2016ARA&A..54..529B} {54, 529}

\bibitem[\protect\citeauthoryear{{Blitz} \& {Spergel}}{{Blitz} \&
  {Spergel}}{1991}]{Bl91}
{Blitz} L.,  {Spergel} D.~N.,  1991, \mn@doi [\apj] {10.1086/170535}, \href
  {https://ui.adsabs.harvard.edu/abs/1991ApJ...379..631B} {379, 631}

\bibitem[\protect\citeauthoryear{{Bovy}}{{Bovy}}{2015}]{galpy}
{Bovy} J.,  2015, \mn@doi [\apjs] {10.1088/0067-0049/216/2/29}, \href
  {https://ui.adsabs.harvard.edu/abs/2015ApJS..216...29B} {216, 29}

\bibitem[\protect\citeauthoryear{{Cabrera-Lavers}, {Hammersley},
  {Gonz{\'a}lez-Fern{\'a}ndez}, {L{\'o}pez-Corredoira}, {Garz{\'o}n}  \&
  {Mahoney}}{{Cabrera-Lavers} et~al.}{2007}]{cabrera-lavers2007}
{Cabrera-Lavers} A.,  {Hammersley} P.~L.,  {Gonz{\'a}lez-Fern{\'a}ndez} C.,
  {L{\'o}pez-Corredoira} M.,  {Garz{\'o}n} F.,   {Mahoney} T.~J.,  2007,
  \mn@doi [\aap] {10.1051/0004-6361:20066185}, \href
  {https://ui.adsabs.harvard.edu/abs/2007A&A...465..825C} {465, 825}

\bibitem[\protect\citeauthoryear{{Cabrera-Lavers},
  {Gonz{\'a}lez-Fern{\'a}ndez}, {Garz{\'o}n}, {Hammersley}  \&
  {L{\'o}pez-Corredoira}}{{Cabrera-Lavers} et~al.}{2008}]{cabrera-lavers2008}
{Cabrera-Lavers} A.,  {Gonz{\'a}lez-Fern{\'a}ndez} C.,  {Garz{\'o}n} F.,
  {Hammersley} P.~L.,   {L{\'o}pez-Corredoira} M.,  2008, \mn@doi [\aap]
  {10.1051/0004-6361:200810720}, \href
  {https://ui.adsabs.harvard.edu/abs/2008A&A...491..781C} {491, 781}

\bibitem[\protect\citeauthoryear{{Ceverino} \& {Klypin}}{{Ceverino} \&
  {Klypin}}{2007}]{Ce07}
{Ceverino} D.,  {Klypin} A.,  2007, \mn@doi [\mnras]
  {10.1111/j.1365-2966.2007.12001.x}, \href
  {https://ui.adsabs.harvard.edu/abs/2007MNRAS.379.1155C} {379, 1155}

\bibitem[\protect\citeauthoryear{{Chiba} \& {Sch{\"o}nrich}}{{Chiba} \&
  {Sch{\"o}nrich}}{2021}]{chiba2021_treering}
{Chiba} R.,  {Sch{\"o}nrich} R.,  2021, \mn@doi [\mnras]
  {10.1093/mnras/stab1094}, \href
  {https://ui.adsabs.harvard.edu/abs/2021MNRAS.505.2412C} {505, 2412}

\bibitem[\protect\citeauthoryear{{Chiba} \& {Sch{\"o}nrich}}{{Chiba} \&
  {Sch{\"o}nrich}}{2022}]{chiba2022}
{Chiba} R.,  {Sch{\"o}nrich} R.,  2022, \mn@doi [\mnras]
  {10.1093/mnras/stac697}, \href
  {https://ui.adsabs.harvard.edu/abs/2022MNRAS.513..768C} {513, 768}

\bibitem[\protect\citeauthoryear{{Chiba}, {Friske}  \& {Sch{\"o}nrich}}{{Chiba}
  et~al.}{2021}]{chiba2021}
{Chiba} R.,  {Friske} J. K.~S.,   {Sch{\"o}nrich} R.,  2021, \mn@doi [\mnras]
  {10.1093/mnras/staa3585}, \href
  {https://ui.adsabs.harvard.edu/abs/2021MNRAS.500.4710C} {500, 4710}

\bibitem[\protect\citeauthoryear{{Churchwell} et~al.,}{{Churchwell}
  et~al.}{2009}]{churchwell2009}
{Churchwell} E.,  et~al., 2009, \mn@doi [\pasp] {10.1086/597811}, \href
  {https://ui.adsabs.harvard.edu/abs/2009PASP..121..213C} {121, 213}

\bibitem[\protect\citeauthoryear{{Collett}, {Dutta}  \& {Evans}}{{Collett}
  et~al.}{1997}]{Co97}
{Collett} J.~L.,  {Dutta} S.~N.,   {Evans} N.~W.,  1997, \mn@doi [\mnras]
  {10.1093/mnras/285.1.49}, \href
  {https://ui.adsabs.harvard.edu/abs/1997MNRAS.285...49C} {285, 49}

\bibitem[\protect\citeauthoryear{{Collier}, {Shlosman}  \& {Heller}}{{Collier}
  et~al.}{2019}]{collier2019}
{Collier} A.,  {Shlosman} I.,   {Heller} C.,  2019, \mn@doi [\mnras]
  {10.1093/mnras/stz2144}, \href
  {https://ui.adsabs.harvard.edu/abs/2019MNRAS.488.5788C} {488, 5788}

\bibitem[\protect\citeauthoryear{{Conroy} et~al.,}{{Conroy}
  et~al.}{2022}]{conroy2022_disk}
{Conroy} C.,  et~al., 2022, arXiv e-prints, \href
  {https://ui.adsabs.harvard.edu/abs/2022arXiv220402989C} {p. arXiv:2204.02989}

\bibitem[\protect\citeauthoryear{{Contopoulos}}{{Contopoulos}}{1980}]{Co80}
{Contopoulos} G.,  1980, \aap, \href
  {https://ui.adsabs.harvard.edu/abs/1980A&A....81..198C} {81, 198}

\bibitem[\protect\citeauthoryear{{Davies}, {Vasiliev}, {Belokurov}, {Evans}  \&
  {Dillamore}}{{Davies} et~al.}{2023a}]{davies2023_ironing}
{Davies} E.~Y.,  {Vasiliev} E.,  {Belokurov} V.,  {Evans} N.~W.,   {Dillamore}
  A.~M.,  2023a, \mn@doi [\mnras] {10.1093/mnras/stac3581}, \href
  {https://ui.adsabs.harvard.edu/abs/2023MNRAS.519..530D} {519, 530}

\bibitem[\protect\citeauthoryear{{Davies}, {Dillamore}, {Vasiliev}  \&
  {Belokurov}}{{Davies} et~al.}{2023b}]{davies2023_bar}
{Davies} E.~Y.,  {Dillamore} A.~M.,  {Vasiliev} E.,   {Belokurov} V.,  2023b,
  \mn@doi [\mnras] {10.1093/mnrasl/slad017}, \href
  {https://ui.adsabs.harvard.edu/abs/2023MNRAS.521L..24D} {521, L24}

\bibitem[\protect\citeauthoryear{{Deason}, {Belokurov}  \& {Evans}}{{Deason}
  et~al.}{2011}]{deason2011_halo}
{Deason} A.~J.,  {Belokurov} V.,   {Evans} N.~W.,  2011, \mn@doi [\mnras]
  {10.1111/j.1365-2966.2011.19237.x}, \href
  {https://ui.adsabs.harvard.edu/abs/2011MNRAS.416.2903D} {416, 2903}

\bibitem[\protect\citeauthoryear{{Debattista} \& {Sellwood}}{{Debattista} \&
  {Sellwood}}{1998}]{Deb98}
{Debattista} V.~P.,  {Sellwood} J.~A.,  1998, \mn@doi [\apjl] {10.1086/311118},
  \href {https://ui.adsabs.harvard.edu/abs/1998ApJ...493L...5D} {493, L5}

\bibitem[\protect\citeauthoryear{{Debattista} \& {Sellwood}}{{Debattista} \&
  {Sellwood}}{2000a}]{De00}
{Debattista} V.~P.,  {Sellwood} J.~A.,  2000a, \mn@doi [\apj] {10.1086/317148},
  \href {https://ui.adsabs.harvard.edu/abs/2000ApJ...543..704D} {543, 704}

\bibitem[\protect\citeauthoryear{{Debattista} \& {Sellwood}}{{Debattista} \&
  {Sellwood}}{2000b}]{debattista2000}
{Debattista} V.~P.,  {Sellwood} J.~A.,  2000b, \mn@doi [\apj] {10.1086/317148},
  \href {https://ui.adsabs.harvard.edu/abs/2000ApJ...543..704D} {543, 704}

\bibitem[\protect\citeauthoryear{{Dehnen}}{{Dehnen}}{1998}]{De98}
{Dehnen} W.,  1998, \mn@doi [\aj] {10.1086/300364}, \href
  {https://ui.adsabs.harvard.edu/abs/1998AJ....115.2384D} {115, 2384}

\bibitem[\protect\citeauthoryear{{Dehnen}}{{Dehnen}}{2000}]{dehnen2000}
{Dehnen} W.,  2000, \mn@doi [\aj] {10.1086/301226}, \href
  {https://ui.adsabs.harvard.edu/abs/2000AJ....119..800D} {119, 800}

\bibitem[\protect\citeauthoryear{{Di Matteo}, {Spite}, {Haywood}, {Bonifacio},
  {G{\'o}mez}, {Spite}  \& {Caffau}}{{Di Matteo} et~al.}{2020}]{DiMatteo2020}
{Di Matteo} P.,  {Spite} M.,  {Haywood} M.,  {Bonifacio} P.,  {G{\'o}mez} A.,
  {Spite} F.,   {Caffau} E.,  2020, \mn@doi [\aap]
  {10.1051/0004-6361/201937016}, \href
  {https://ui.adsabs.harvard.edu/abs/2020A&A...636A.115D} {636, A115}

\bibitem[\protect\citeauthoryear{{Dong-P{\'a}ez}, {Vasiliev}  \&
  {Evans}}{{Dong-P{\'a}ez} et~al.}{2022}]{dong-paez2022}
{Dong-P{\'a}ez} C.~A.,  {Vasiliev} E.,   {Evans} N.~W.,  2022, \mn@doi [\mnras]
  {10.1093/mnras/stab3361}, \href
  {https://ui.adsabs.harvard.edu/abs/2022MNRAS.510..230D} {510, 230}

\bibitem[\protect\citeauthoryear{{Faccioli}, {Smith}, {Yuan}, {Zhang}, {Liu},
  {Zhao}  \& {Yao}}{{Faccioli} et~al.}{2014}]{faccioli2014_halo}
{Faccioli} L.,  {Smith} M.~C.,  {Yuan} H.~B.,  {Zhang} H.~H.,  {Liu} X.~W.,
  {Zhao} H.~B.,   {Yao} J.~S.,  2014, \mn@doi [\apj]
  {10.1088/0004-637X/788/2/105}, \href
  {https://ui.adsabs.harvard.edu/abs/2014ApJ...788..105F} {788, 105}

\bibitem[\protect\citeauthoryear{{Feuillet}, {Sahlholdt}, {Feltzing}  \&
  {Casagrande}}{{Feuillet} et~al.}{2021}]{feuillet2021}
{Feuillet} D.~K.,  {Sahlholdt} C.~L.,  {Feltzing} S.,   {Casagrande} L.,  2021,
  \mn@doi [\mnras] {10.1093/mnras/stab2614}, \href
  {https://ui.adsabs.harvard.edu/abs/2021MNRAS.508.1489F} {508, 1489}

\bibitem[\protect\citeauthoryear{{Fillmore} \& {Goldreich}}{{Fillmore} \&
  {Goldreich}}{1984}]{filmore1984}
{Fillmore} J.~A.,  {Goldreich} P.,  1984, \mn@doi [\apj] {10.1086/162070},
  \href {https://ui.adsabs.harvard.edu/abs/1984ApJ...281....1F} {281, 1}

\bibitem[\protect\citeauthoryear{{Fragkoudi} et~al.,}{{Fragkoudi}
  et~al.}{2019}]{Fr19}
{Fragkoudi} F.,  et~al., 2019, \mn@doi [\mnras] {10.1093/mnras/stz1875}, \href
  {https://ui.adsabs.harvard.edu/abs/2019MNRAS.488.3324F} {488, 3324}

\bibitem[\protect\citeauthoryear{{Fux}}{{Fux}}{1999}]{Fu99}
{Fux} R.,  1999, \mn@doi [\aap] {10.48550/arXiv.astro-ph/9903154}, \href
  {https://ui.adsabs.harvard.edu/abs/1999A&A...345..787F} {345, 787}

\bibitem[\protect\citeauthoryear{{Gaia Collaboration} et~al.,}{{Gaia
  Collaboration} et~al.}{2016}]{gaia}
{Gaia Collaboration} et~al., 2016, \mn@doi [\aap]
  {10.1051/0004-6361/201629272}, \href
  {https://ui.adsabs.harvard.edu/abs/2016A&A...595A...1G} {595, A1}

\bibitem[\protect\citeauthoryear{{Gaia Collaboration} et~al.,}{{Gaia
  Collaboration} et~al.}{2023a}]{gaia_dr3}
{Gaia Collaboration} et~al., 2023a, \mn@doi [\aap]
  {10.1051/0004-6361/202243940}, \href
  {https://ui.adsabs.harvard.edu/abs/2023A&A...674A...1G} {674, A1}

\bibitem[\protect\citeauthoryear{{Gaia Collaboration} et~al.,}{{Gaia
  Collaboration} et~al.}{2023b}]{gaia_dr3_disc}
{Gaia Collaboration} et~al., 2023b, \mn@doi [\aap]
  {10.1051/0004-6361/202243797}, \href
  {https://ui.adsabs.harvard.edu/abs/2023A&A...674A..37G} {674, A37}

\bibitem[\protect\citeauthoryear{{Gonz{\'a}lez-Fern{\'a}ndez},
  {L{\'o}pez-Corredoira}, {Am{\^o}res}, {Minniti}, {Lucas}  \&
  {Toledo}}{{Gonz{\'a}lez-Fern{\'a}ndez} et~al.}{2012}]{Go12}
{Gonz{\'a}lez-Fern{\'a}ndez} C.,  {L{\'o}pez-Corredoira} M.,  {Am{\^o}res}
  E.~B.,  {Minniti} D.,  {Lucas} P.,   {Toledo} I.,  2012, \mn@doi [\aap]
  {10.1051/0004-6361/201219756}, \href
  {https://ui.adsabs.harvard.edu/abs/2012A&A...546A.107G} {546, A107}

\bibitem[\protect\citeauthoryear{{Gurvich} et~al.,}{{Gurvich}
  et~al.}{2023}]{Gurvich2023}
{Gurvich} A.~B.,  et~al., 2023, \mn@doi [\mnras] {10.1093/mnras/stac3712},
  \href {https://ui.adsabs.harvard.edu/abs/2023MNRAS.519.2598G} {519, 2598}

\bibitem[\protect\citeauthoryear{{Hammersley}, {Garzon}, {Mahoney}  \&
  {Calbet}}{{Hammersley} et~al.}{1994}]{hammersley1994}
{Hammersley} P.~L.,  {Garzon} F.,  {Mahoney} T.,   {Calbet} X.,  1994, \mn@doi
  [\mnras] {10.1093/mnras/269.3.753}, \href
  {https://ui.adsabs.harvard.edu/abs/1994MNRAS.269..753H} {269, 753}

\bibitem[\protect\citeauthoryear{{Hammersley}, {Garz{\'o}n}, {Mahoney},
  {L{\'o}pez-Corredoira}  \& {Torres}}{{Hammersley}
  et~al.}{2000}]{hammersley2000}
{Hammersley} P.~L.,  {Garz{\'o}n} F.,  {Mahoney} T.~J.,  {L{\'o}pez-Corredoira}
  M.,   {Torres} M.~A.~P.,  2000, \mn@doi [\mnras]
  {10.1046/j.1365-8711.2000.03858.x}, \href
  {https://ui.adsabs.harvard.edu/abs/2000MNRAS.317L..45H} {317, L45}

\bibitem[\protect\citeauthoryear{{Hattori}, {Erkal}  \& {Sanders}}{{Hattori}
  et~al.}{2016}]{Ha16}
{Hattori} K.,  {Erkal} D.,   {Sanders} J.~L.,  2016, \mn@doi [\mnras]
  {10.1093/mnras/stw1006}, \href
  {https://ui.adsabs.harvard.edu/abs/2016MNRAS.460..497H} {460, 497}

\bibitem[\protect\citeauthoryear{{Helmi}, {Babusiaux}, {Koppelman}, {Massari},
  {Veljanoski}  \& {Brown}}{{Helmi} et~al.}{2018}]{helmi2018}
{Helmi} A.,  {Babusiaux} C.,  {Koppelman} H.~H.,  {Massari} D.,  {Veljanoski}
  J.,   {Brown} A. G.~A.,  2018, \mn@doi [\nat] {10.1038/s41586-018-0625-x},
  \href {https://ui.adsabs.harvard.edu/abs/2018Natur.563...85H} {563, 85}

\bibitem[\protect\citeauthoryear{{Hernquist} \& {Weinberg}}{{Hernquist} \&
  {Weinberg}}{1992}]{hernquist1992}
{Hernquist} L.,  {Weinberg} M.~D.,  1992, \mn@doi [\apj] {10.1086/171975},
  \href {https://ui.adsabs.harvard.edu/abs/1992ApJ...400...80H} {400, 80}

\bibitem[\protect\citeauthoryear{{Hopkins} et~al.,}{{Hopkins}
  et~al.}{2023}]{Hopkins2023}
{Hopkins} P.~F.,  et~al., 2023, \mn@doi [\mnras] {10.1093/mnras/stad1902},
  \href {https://ui.adsabs.harvard.edu/abs/2023MNRAS.tmp.1847H} {}

\bibitem[\protect\citeauthoryear{{Hunt}, {Hong}, {Bovy}, {Kawata}  \&
  {Grand}}{{Hunt} et~al.}{2018}]{Hu18}
{Hunt} J. A.~S.,  {Hong} J.,  {Bovy} J.,  {Kawata} D.,   {Grand} R. J.~J.,
  2018, \mn@doi [\mnras] {10.1093/mnras/sty2532}, \href
  {https://ui.adsabs.harvard.edu/abs/2018MNRAS.481.3794H} {481, 3794}

\bibitem[\protect\citeauthoryear{{Kalnajs}}{{Kalnajs}}{1991}]{Ka91}
{Kalnajs} A.~J.,  1991, in {Sundelius} B.,  ed., Dynamics of Disc Galaxies.
  p.~323

\bibitem[\protect\citeauthoryear{{Katz} et~al.,}{{Katz}
  et~al.}{2023}]{gaia_rvs}
{Katz} D.,  et~al., 2023, \mn@doi [\aap] {10.1051/0004-6361/202244220}, \href
  {https://ui.adsabs.harvard.edu/abs/2023A&A...674A...5K} {674, A5}

\bibitem[\protect\citeauthoryear{{Khoperskov}, {Gerhard}, {Di Matteo},
  {Haywood}, {Katz}, {Khrapov}, {Khoperskov}  \& {Arnaboldi}}{{Khoperskov}
  et~al.}{2020}]{khoperskov2020}
{Khoperskov} S.,  {Gerhard} O.,  {Di Matteo} P.,  {Haywood} M.,  {Katz} D.,
  {Khrapov} S.,  {Khoperskov} A.,   {Arnaboldi} M.,  2020, \mn@doi [\aap]
  {10.1051/0004-6361/201936645}, \href
  {https://ui.adsabs.harvard.edu/abs/2020A&A...634L...8K} {634, L8}

\bibitem[\protect\citeauthoryear{{Kruijssen} et~al.,}{{Kruijssen}
  et~al.}{2020}]{kruijssen2020}
{Kruijssen} J.~M.~D.,  et~al., 2020, \mn@doi [\mnras] {10.1093/mnras/staa2452},
  \href {https://ui.adsabs.harvard.edu/abs/2020MNRAS.498.2472K} {498, 2472}

\bibitem[\protect\citeauthoryear{{Li}, {Shen}, {Gerhard}  \& {Clarke}}{{Li}
  et~al.}{2022}]{Li22}
{Li} Z.,  {Shen} J.,  {Gerhard} O.,   {Clarke} J.~P.,  2022, \mn@doi [\apj]
  {10.3847/1538-4357/ac3823}, \href
  {https://ui.adsabs.harvard.edu/abs/2022ApJ...925...71L} {925, 71}

\bibitem[\protect\citeauthoryear{{Lucey}, {Pearson}, {Hunt}, {Hawkins}, {Ness},
  {Petersen}, {Price-Whelan}  \& {Weinberg}}{{Lucey} et~al.}{2023}]{lucey2022}
{Lucey} M.,  {Pearson} S.,  {Hunt} J. A.~S.,  {Hawkins} K.,  {Ness} M.,
  {Petersen} M.~S.,  {Price-Whelan} A.~M.,   {Weinberg} M.~D.,  2023, \mn@doi
  [\mnras] {10.1093/mnras/stad406}, \href
  {https://ui.adsabs.harvard.edu/abs/2023MNRAS.520.4779L} {520, 4779}

\bibitem[\protect\citeauthoryear{{Lynden-Bell}}{{Lynden-Bell}}{1973}]{Ly73}
{Lynden-Bell} D.,  1973, in {Contopoulos} G.,  {Henon} M.,   {Lynden-Bell} D.,
  eds, Saas-Fee Advanced Course 3: Dynamical Structure and Evolution of Stellar
  Systems. p.~91

\bibitem[\protect\citeauthoryear{{Lynden-Bell} \& {Kalnajs}}{{Lynden-Bell} \&
  {Kalnajs}}{1972}]{LBK}
{Lynden-Bell} D.,  {Kalnajs} A.~J.,  1972, \mn@doi [\mnras]
  {10.1093/mnras/157.1.1}, \href
  {https://ui.adsabs.harvard.edu/abs/1972MNRAS.157....1L} {157, 1}

\bibitem[\protect\citeauthoryear{{Majewski} et~al.,}{{Majewski}
  et~al.}{2017}]{apogee}
{Majewski} S.~R.,  et~al., 2017, \mn@doi [\aj] {10.3847/1538-3881/aa784d},
  \href {https://ui.adsabs.harvard.edu/abs/2017AJ....154...94M} {154, 94}

\bibitem[\protect\citeauthoryear{{Mardini}, {Frebel}, {Chiti}, {Meiron},
  {Brauer}  \& {Ou}}{{Mardini} et~al.}{2022}]{Mardini2022}
{Mardini} M.~K.,  {Frebel} A.,  {Chiti} A.,  {Meiron} Y.,  {Brauer} K.~V.,
  {Ou} X.,  2022, \mn@doi [\apj] {10.3847/1538-4357/ac8102}, \href
  {https://ui.adsabs.harvard.edu/abs/2022ApJ...936...78M} {936, 78}

\bibitem[\protect\citeauthoryear{{Martinez-Valpuesta} \&
  {Gerhard}}{{Martinez-Valpuesta} \& {Gerhard}}{2011}]{martinez-valpuesta2011}
{Martinez-Valpuesta} I.,  {Gerhard} O.,  2011, \mn@doi [\apjl]
  {10.1088/2041-8205/734/1/L20}, \href
  {https://ui.adsabs.harvard.edu/abs/2011ApJ...734L..20M} {734, L20}

\bibitem[\protect\citeauthoryear{{Molloy}, {Smith}, {Shen}  \&
  {Evans}}{{Molloy} et~al.}{2015}]{Mo15}
{Molloy} M.,  {Smith} M.~C.,  {Shen} J.,   {Evans} N.~W.,  2015, \mn@doi [\apj]
  {10.1088/0004-637X/804/2/80}, \href
  {https://ui.adsabs.harvard.edu/abs/2015ApJ...804...80M} {804, 80}

\bibitem[\protect\citeauthoryear{{Monari}, {Famaey}, {Siebert}, {Grand},
  {Kawata}  \& {Boily}}{{Monari} et~al.}{2016}]{monari2016}
{Monari} G.,  {Famaey} B.,  {Siebert} A.,  {Grand} R. J.~J.,  {Kawata} D.,
  {Boily} C.,  2016, \mn@doi [\mnras] {10.1093/mnras/stw1564}, \href
  {https://ui.adsabs.harvard.edu/abs/2016MNRAS.461.3835M} {461, 3835}

\bibitem[\protect\citeauthoryear{{Monari}, {Famaey}, {Siebert}, {Wegg}  \&
  {Gerhard}}{{Monari} et~al.}{2019}]{Mo19}
{Monari} G.,  {Famaey} B.,  {Siebert} A.,  {Wegg} C.,   {Gerhard} O.,  2019,
  \mn@doi [\aap] {10.1051/0004-6361/201834820}, \href
  {https://ui.adsabs.harvard.edu/abs/2019A&A...626A..41M} {626, A41}

\bibitem[\protect\citeauthoryear{{Moreno}, {Pichardo}  \& {Schuster}}{{Moreno}
  et~al.}{2015a}]{moreno2015}
{Moreno} E.,  {Pichardo} B.,   {Schuster} W.~J.,  2015a, \mn@doi [\mnras]
  {10.1093/mnras/stv962}, \href
  {https://ui.adsabs.harvard.edu/abs/2015MNRAS.451..705M} {451, 705}

\bibitem[\protect\citeauthoryear{{Moreno}, {Pichardo}  \& {Schuster}}{{Moreno}
  et~al.}{2015b}]{Mor15}
{Moreno} E.,  {Pichardo} B.,   {Schuster} W.~J.,  2015b, \mn@doi [\mnras]
  {10.1093/mnras/stv962}, \href
  {https://ui.adsabs.harvard.edu/abs/2015MNRAS.451..705M} {451, 705}

\bibitem[\protect\citeauthoryear{{Myeong}, {Evans}, {Belokurov}, {Sanders}  \&
  {Koposov}}{{Myeong} et~al.}{2018}]{My18}
{Myeong} G.~C.,  {Evans} N.~W.,  {Belokurov} V.,  {Sanders} J.~L.,   {Koposov}
  S.~E.,  2018, \mn@doi [\apjl] {10.3847/2041-8213/aab613}, \href
  {https://ui.adsabs.harvard.edu/abs/2018ApJ...856L..26M} {856, L26}

\bibitem[\protect\citeauthoryear{{Myeong}, {Belokurov}, {Aguado}, {Evans},
  {Caldwell}  \& {Bradley}}{{Myeong} et~al.}{2022}]{myeong2022_eccentric}
{Myeong} G.~C.,  {Belokurov} V.,  {Aguado} D.~S.,  {Evans} N.~W.,  {Caldwell}
  N.,   {Bradley} J.,  2022, \mn@doi [\apj] {10.3847/1538-4357/ac8d68}, \href
  {https://ui.adsabs.harvard.edu/abs/2022ApJ...938...21M} {938, 21}

\bibitem[\protect\citeauthoryear{{Naidu}, {Conroy}, {Bonaca}, {Johnson},
  {Ting}, {Caldwell}, {Zaritsky}  \& {Cargile}}{{Naidu}
  et~al.}{2020}]{naidu2020}
{Naidu} R.~P.,  {Conroy} C.,  {Bonaca} A.,  {Johnson} B.~D.,  {Ting} Y.-S.,
  {Caldwell} N.,  {Zaritsky} D.,   {Cargile} P.~A.,  2020, \mn@doi [\apj]
  {10.3847/1538-4357/abaef4}, \href
  {https://ui.adsabs.harvard.edu/abs/2020ApJ...901...48N} {901, 48}

\bibitem[\protect\citeauthoryear{{Pila-D{\'\i}ez}, {de Jong}, {Kuijken}, {van
  der Burg}  \& {Hoekstra}}{{Pila-D{\'\i}ez} et~al.}{2015}]{pila-diez2015}
{Pila-D{\'\i}ez} B.,  {de Jong} J.~T.~A.,  {Kuijken} K.,  {van der Burg}
  R.~F.~J.,   {Hoekstra} H.,  2015, \mn@doi [\aap]
  {10.1051/0004-6361/201425457}, \href
  {https://ui.adsabs.harvard.edu/abs/2015A&A...579A..38P} {579, A38}

\bibitem[\protect\citeauthoryear{{Portail}, {Gerhard}, {Wegg}  \&
  {Ness}}{{Portail} et~al.}{2017}]{Po17}
{Portail} M.,  {Gerhard} O.,  {Wegg} C.,   {Ness} M.,  2017, \mn@doi [\mnras]
  {10.1093/mnras/stw2819}, \href
  {https://ui.adsabs.harvard.edu/abs/2017MNRAS.465.1621P} {465, 1621}

\bibitem[\protect\citeauthoryear{{Posti}, {Binney}, {Nipoti}  \&
  {Ciotti}}{{Posti} et~al.}{2015}]{posti2015}
{Posti} L.,  {Binney} J.,  {Nipoti} C.,   {Ciotti} L.,  2015, \mn@doi [\mnras]
  {10.1093/mnras/stu2608}, \href
  {https://ui.adsabs.harvard.edu/abs/2015MNRAS.447.3060P} {447, 3060}

\bibitem[\protect\citeauthoryear{{Rix} et~al.,}{{Rix} et~al.}{2022}]{Rix2022}
{Rix} H.-W.,  et~al., 2022, \mn@doi [\apj] {10.3847/1538-4357/ac9e01}, \href
  {https://ui.adsabs.harvard.edu/abs/2022ApJ...941...45R} {941, 45}

\bibitem[\protect\citeauthoryear{{Rosas-Guevara} et~al.,}{{Rosas-Guevara}
  et~al.}{2022}]{rosas-guevara2022}
{Rosas-Guevara} Y.,  et~al., 2022, \mn@doi [\mnras] {10.1093/mnras/stac816},
  \href {https://ui.adsabs.harvard.edu/abs/2022MNRAS.512.5339R} {512, 5339}

\bibitem[\protect\citeauthoryear{{Saha}, {Martinez-Valpuesta}  \&
  {Gerhard}}{{Saha} et~al.}{2012}]{saha2012}
{Saha} K.,  {Martinez-Valpuesta} I.,   {Gerhard} O.,  2012, \mn@doi [\mnras]
  {10.1111/j.1365-2966.2011.20307.x}, \href
  {https://ui.adsabs.harvard.edu/abs/2012MNRAS.421..333S} {421, 333}

\bibitem[\protect\citeauthoryear{{Saha}, {Gerhard}  \&
  {Martinez-Valpuesta}}{{Saha} et~al.}{2016}]{saha2016}
{Saha} K.,  {Gerhard} O.,   {Martinez-Valpuesta} I.,  2016, \mn@doi [\aap]
  {10.1051/0004-6361/201527566}, \href
  {https://ui.adsabs.harvard.edu/abs/2016A&A...588A..42S} {588, A42}

\bibitem[\protect\citeauthoryear{{Sanders} \& {Binney}}{{Sanders} \&
  {Binney}}{2016}]{sanders2016}
{Sanders} J.~L.,  {Binney} J.,  2016, \mn@doi [\mnras] {10.1093/mnras/stw106},
  \href {https://ui.adsabs.harvard.edu/abs/2016MNRAS.457.2107S} {457, 2107}

\bibitem[\protect\citeauthoryear{{Sanders}, {Smith}  \& {Evans}}{{Sanders}
  et~al.}{2019}]{Sa19}
{Sanders} J.~L.,  {Smith} L.,   {Evans} N.~W.,  2019, \mn@doi [\mnras]
  {10.1093/mnras/stz1827}, \href
  {https://ui.adsabs.harvard.edu/abs/2019MNRAS.488.4552S} {488, 4552}

\bibitem[\protect\citeauthoryear{{Sanderson} \& {Helmi}}{{Sanderson} \&
  {Helmi}}{2013}]{sanderson2013}
{Sanderson} R.~E.,  {Helmi} A.,  2013, \mn@doi [\mnras]
  {10.1093/mnras/stt1307}, \href
  {https://ui.adsabs.harvard.edu/abs/2013MNRAS.435..378S} {435, 378}

\bibitem[\protect\citeauthoryear{{Schuster}, {Moreno}  \&
  {Fern{\'a}ndez-Trincado}}{{Schuster} et~al.}{2019}]{Sc19}
{Schuster} W.~J.,  {Moreno} E.,   {Fern{\'a}ndez-Trincado} J.~G.,  2019, in
  {McQuinn} K. B.~W.,  {Stierwalt} S.,  eds,  Proc. IAU Symp. Vol. 344, Dwarf
  Galaxies: From the Deep Universe to the Present. pp 134--138 (\mn@eprint
  {arXiv} {1811.09827}), \mn@doi{10.1017/S174392131800683X}

\bibitem[\protect\citeauthoryear{{Sestito} et~al.,}{{Sestito}
  et~al.}{2019}]{sestito2019}
{Sestito} F.,  et~al., 2019, \mn@doi [\mnras] {10.1093/mnras/stz043}, \href
  {https://ui.adsabs.harvard.edu/abs/2019MNRAS.484.2166S} {484, 2166}

\bibitem[\protect\citeauthoryear{{Shu}}{{Shu}}{1991}]{Shu}
{Shu} F.,  1991, {Physics of Astrophysics, Vol. II: Gas Dynamics}

\bibitem[\protect\citeauthoryear{{Silva}, {Schuster}  \& {Contreras}}{{Silva}
  et~al.}{2012}]{Si12}
{Silva} J.~S.,  {Schuster} W.~J.,   {Contreras} M.~E.,  2012, \rmxaa, \href
  {https://ui.adsabs.harvard.edu/abs/2012RMxAA..48..109S} {48, 109}

\bibitem[\protect\citeauthoryear{{Simion}, {Belokurov}, {Irwin}, {Koposov},
  {Gonzalez-Fernandez}, {Robin}, {Shen}  \& {Li}}{{Simion}
  et~al.}{2017}]{Simion2017}
{Simion} I.~T.,  {Belokurov} V.,  {Irwin} M.,  {Koposov} S.~E.,
  {Gonzalez-Fernandez} C.,  {Robin} A.~C.,  {Shen} J.,   {Li} Z.~Y.,  2017,
  \mn@doi [\mnras] {10.1093/mnras/stx1832}, \href
  {https://ui.adsabs.harvard.edu/abs/2017MNRAS.471.4323S} {471, 4323}

\bibitem[\protect\citeauthoryear{{Sormani}, {Binney}  \& {Magorrian}}{{Sormani}
  et~al.}{2015}]{So15}
{Sormani} M.~C.,  {Binney} J.,   {Magorrian} J.,  2015, \mn@doi [\mnras]
  {10.1093/mnras/stv2067}, \href
  {https://ui.adsabs.harvard.edu/abs/2015MNRAS.454.1818S} {454, 1818}

\bibitem[\protect\citeauthoryear{{Stanek}, {Mateo}, {Udalski}, {Szymanski},
  {Kaluzny}  \& {Kubiak}}{{Stanek} et~al.}{1994}]{St94}
{Stanek} K.~Z.,  {Mateo} M.,  {Udalski} A.,  {Szymanski} M.,  {Kaluzny} J.,
  {Kubiak} M.,  1994, \mn@doi [\apjl] {10.1086/187416}, \href
  {https://ui.adsabs.harvard.edu/abs/1994ApJ...429L..73S} {429, L73}

\bibitem[\protect\citeauthoryear{{Stanek}, {Udalski}, {Szyma{\'N}ski},
  {Ka{\L}u{\.Z}ny}, {Kubiak}, {Mateo}  \& {Krzemi{\'N}ski}}{{Stanek}
  et~al.}{1997}]{St97}
{Stanek} K.~Z.,  {Udalski} A.,  {Szyma{\'N}ski} M.,  {Ka{\L}u{\.Z}ny} J.,
  {Kubiak} Z.~M.,  {Mateo} M.,   {Krzemi{\'N}ski} W.,  1997, \mn@doi [\apj]
  {10.1086/303702}, \href
  {https://ui.adsabs.harvard.edu/abs/1997ApJ...477..163S} {477, 163}

\bibitem[\protect\citeauthoryear{{Tremaine} \& {Weinberg}}{{Tremaine} \&
  {Weinberg}}{1984}]{Tr84}
{Tremaine} S.,  {Weinberg} M.~D.,  1984, \mn@doi [\apjl] {10.1086/184292},
  \href {https://ui.adsabs.harvard.edu/abs/1984ApJ...282L...5T} {282, L5}

\bibitem[\protect\citeauthoryear{{Trick}}{{Trick}}{2022}]{Tr22}
{Trick} W.~H.,  2022, \mn@doi [\mnras] {10.1093/mnras/stab2866}, \href
  {https://ui.adsabs.harvard.edu/abs/2022MNRAS.509..844T} {509, 844}

\bibitem[\protect\citeauthoryear{{Trick}, {Fragkoudi}, {Hunt}, {Mackereth}  \&
  {White}}{{Trick} et~al.}{2021}]{trick2021}
{Trick} W.~H.,  {Fragkoudi} F.,  {Hunt} J. A.~S.,  {Mackereth} J.~T.,   {White}
  S. D.~M.,  2021, \mn@doi [\mnras] {10.1093/mnras/staa3317}, \href
  {https://ui.adsabs.harvard.edu/abs/2021MNRAS.500.2645T} {500, 2645}

\bibitem[\protect\citeauthoryear{{Vasiliev}}{{Vasiliev}}{2019}]{agama}
{Vasiliev} E.,  2019, \mn@doi [\mnras] {10.1093/mnras/sty2672}, \href
  {https://ui.adsabs.harvard.edu/abs/2019MNRAS.482.1525V} {482, 1525}

\bibitem[\protect\citeauthoryear{{Wang}, {Mao}, {Long}  \& {Shen}}{{Wang}
  et~al.}{2013}]{wang2013}
{Wang} Y.,  {Mao} S.,  {Long} R.~J.,   {Shen} J.,  2013, \mn@doi [\mnras]
  {10.1093/mnras/stt1537}, \href
  {https://ui.adsabs.harvard.edu/abs/2013MNRAS.435.3437W} {435, 3437}

\bibitem[\protect\citeauthoryear{{Watkins} et~al.,}{{Watkins}
  et~al.}{2009}]{Wa09}
{Watkins} L.~L.,  et~al., 2009, \mn@doi [\mnras]
  {10.1111/j.1365-2966.2009.15242.x}, \href
  {https://ui.adsabs.harvard.edu/abs/2009MNRAS.398.1757W} {398, 1757}

\bibitem[\protect\citeauthoryear{{Wegg}, {Gerhard}  \& {Portail}}{{Wegg}
  et~al.}{2015}]{wegg2015}
{Wegg} C.,  {Gerhard} O.,   {Portail} M.,  2015, \mn@doi [\mnras]
  {10.1093/mnras/stv745}, \href
  {https://ui.adsabs.harvard.edu/abs/2015MNRAS.450.4050W} {450, 4050}

\bibitem[\protect\citeauthoryear{{Weiland} et~al.,}{{Weiland}
  et~al.}{1994}]{We94}
{Weiland} J.~L.,  et~al., 1994, \mn@doi [\apjl] {10.1086/187315}, \href
  {https://ui.adsabs.harvard.edu/abs/1994ApJ...425L..81W} {425, L81}

\bibitem[\protect\citeauthoryear{{Weinberg}}{{Weinberg}}{1985}]{weinberg1985}
{Weinberg} M.~D.,  1985, \mn@doi [\mnras] {10.1093/mnras/213.3.451}, \href
  {https://ui.adsabs.harvard.edu/abs/1985MNRAS.213..451W} {213, 451}

\bibitem[\protect\citeauthoryear{{Weinberg} \& {Katz}}{{Weinberg} \&
  {Katz}}{2002}]{We02}
{Weinberg} M.~D.,  {Katz} N.,  2002, \mn@doi [\apj] {10.1086/343847}, \href
  {https://ui.adsabs.harvard.edu/abs/2002ApJ...580..627W} {580, 627}

\bibitem[\protect\citeauthoryear{{Wheeler}, {Abril-Cabezas}, {Trick},
  {Fragkoudi}  \& {Ness}}{{Wheeler} et~al.}{2022}]{wheeler2022}
{Wheeler} A.,  {Abril-Cabezas} I.,  {Trick} W.~H.,  {Fragkoudi} F.,   {Ness}
  M.,  2022, \mn@doi [\apj] {10.3847/1538-4357/ac7da0}, \href
  {https://ui.adsabs.harvard.edu/abs/2022ApJ...935...28W} {935, 28}

\bibitem[\protect\citeauthoryear{{Whitelock}}{{Whitelock}}{1992}]{Wh92}
{Whitelock} P.,  1992, in {Warner} B.,  ed.,  Astronomical Society of the
  Pacific Conference Series Vol. 30, Variable Stars and Galaxies, in honor of
  M. W. Feast on his retirement. p.~11

\bibitem[\protect\citeauthoryear{{Williams} \& {Evans}}{{Williams} \&
  {Evans}}{2015}]{Wi15}
{Williams} A.~A.,  {Evans} N.~W.,  2015, \mn@doi [\mnras]
  {10.1093/mnras/stv096}, \href
  {https://ui.adsabs.harvard.edu/abs/2015MNRAS.448.1360W} {448, 1360}

\bibitem[\protect\citeauthoryear{{de Vaucouleurs}}{{de
  Vaucouleurs}}{1964}]{Va64}
{de Vaucouleurs} G.,  1964, in {Kerr} F.~J.,  ed.,  Proc. IAU Symp. Vol. 20,
  The Galaxy and the Magellanic Clouds. p.~195

\bibitem[\protect\citeauthoryear{{de Zeeuw}}{{de Zeeuw}}{1985}]{dezeeuw1985}
{de Zeeuw} T.,  1985, \mn@doi [\mnras] {10.1093/mnras/216.2.273}, \href
  {https://ui.adsabs.harvard.edu/abs/1985MNRAS.216..273D} {216, 273}

\makeatother
\end{thebibliography}



\appendix





\bsp	
\label{lastpage}
\end{document}